\documentclass[prl,aps,showkeys,showpacs,superscriptaddress,twocolumn,%
amsmath,amssymb]{revtex4} 

\usepackage{graphics}
\usepackage{epsfig}
\usepackage{epstopdf}
\usepackage{amsmath}
\usepackage{setspace}
\usepackage[bookmarks=false,linkcolor=blue,urlcolor=black,colorlinks,citecolor=blue]{hyperref}

\begin{document}
\title{Microscopic origin of frictional rheology in dense suspensions:\\  correlations in force space}
\author{Jetin E. Thomas} 
\email{jethomas@brandeis.edu}
\affiliation{Martin Fisher School of Physics, Brandeis University, Waltham, MA 02454, USA}
\author{Kabir Ramola} 
\email{kramola@brandeis.edu}
\affiliation{Martin Fisher School of Physics, Brandeis University, Waltham, MA 02454, USA}
\author{Abhinendra Singh}
\email{abhinendra@uchicago.edu}
\affiliation{Benjamin Levich Institute, CUNY City College of New York, New York, NY 10031, USA}
\author{Romain Mari}
\email{romain.mari@univ-grenoble-alpes.fr}
\affiliation{Universit\'e Grenoble Alpes, CNRS, LIPhy, 38000 Grenoble, France}
\author{Jeffrey F. Morris}
\email{morris@ccny.cuny.edu}
\affiliation{Benjamin Levich Institute, CUNY City College of New York, New York, NY 10031, USA}
\affiliation{Department of Chemical Engineering, CUNY City College of New York, New York, NY 10031}
\author{Bulbul Chakraborty}
\email{bulbul@brandeis.edu}
\affiliation{Martin Fisher School of Physics, Brandeis University, Waltham, MA 02454, USA}

\date{\today} 


\begin{abstract}
We develop a statistical framework for the rheology of dense, non-Brownian suspensions,
based on correlations in a  space representing forces, which is dual to position space. Working with the ensemble of steady state configurations obtained from simulations of suspensions in two dimensions, we find that the anisotropy of the pair correlation function in force space  changes with confining shear stress ($\sigma_{xy}$) and packing fraction ($\phi$). Using these microscopic correlations, we build a statistical theory for the macroscopic friction coefficient: the anisotropy of the stress tensor,  $\mu = \sigma_{xy}/P$. We find that  $\mu$ decreases (i) as $\phi$ is increased and (ii) as $\sigma_{xy}$ is increased. Using a new constitutive relation between  $\mu$ and viscosity for dense suspensions that generalizes the rate-independent one, we show that our theory predicts a Discontinuous Shear Thickening (DST) flow diagram that is in good agreement with numerical simulations, and the qualitative features of $\mu$ that lead to the generic flow diagram of a DST fluid observed in experiments.
\end{abstract}

\maketitle

Dense suspensions of frictional grains in a fluid often display an increase in viscosity $\eta = \sigma_{xy}/\dot{\gamma}$ (thickening) as the confining shear stress ($\sigma_{xy}$) or strain rate ($\dot{\gamma}$) are increased.
At a critical density dependent shear rate $\dot{\gamma}$, the viscosity increases abruptly: a phenomenon termed Discontinuous Shear Thickening (DST). In stress-controlled protocols, $\eta \sim \sigma_{xy}$ marks the DST boundary~\cite{Mewis:2012zl,Brown:2014uq}.
Experiments have also observed interesting features in other components of the stress tensor such as the first normal stress difference, $N_1 = \sigma_{xx}-\sigma_{yy}$ close to the DST  regime \cite{royer2016rheological}.  A mean-field theory~\cite{wyart2014discontinuous,cates2014granulation},  based on  an increase in the fraction of close interactions becoming frictional (rather than lubricated) with increasing shear stress, has been extremely successful at predicting the flow curves and the DST flow diagram in the space of packing fraction, $\phi$ and shear stress or strain rate~\cite{mari2014shear,singh2018constitutive}. The physical picture of lubricated layers between grains giving way to frictional contacts when the imposed $\sigma_{xy}$ exceeds a critical value set by a repulsive force \cite{wyart2014discontinuous} provides a consistent theory  of DST~\cite{singh2018constitutive}, shear jamming fronts \cite{han2017constitutive} and instabilities of the shear-thickened state \cite{hermes2016unsteady}.

\begin{figure}[t!]
\hspace*{-0.4cm}
\includegraphics[width=1.1\linewidth]{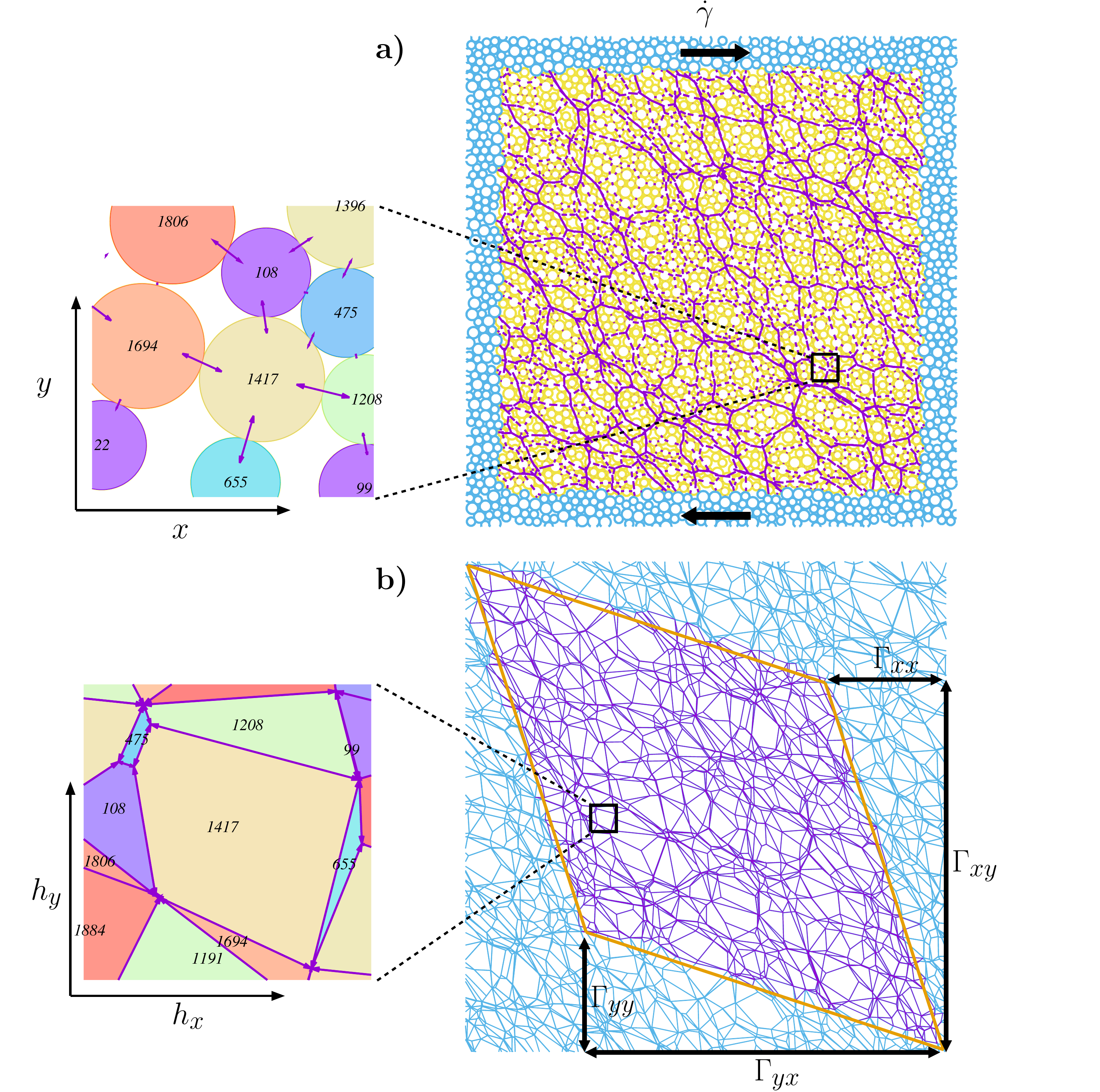}
\caption{(Color online) {\bf a)} A snapshot of a sheared suspension of $2000$ soft frictional disks. The lines represent the pairwise (lubricated and frictional contact) force vectors between the individual grains. {\bf b)} The force tiling associated with this flowing dense suspension. The bonds correspond to the pairwise forces, with larger polygons representing grains with higher stress. The vertices of the tiling represent height vectors $\vec{h} = (h_x, h_y)$, whose difference provides the pairwise force at each bond.
$\vec{\Gamma}_x = (\Gamma_{xx},\Gamma_{xy} )$ and $\vec{\Gamma}_y = (\Gamma_{yx},\Gamma_{yy} )$ represent the sum of forces in the $x$ and $y$ directions respectively. 
The light blue regions represent periodic copies of the system.} 
\label{bodyforce_tile_figure}
\end{figure}

Although several features relating to the flow of dense suspensions can be well explained within this  mean-field theory, the nature of the microscopic correlations underlying this transition remains far from clear~\cite{mari2014shear}.  Conventional measures such as the pair correlation function do not exhibit pronounced  changes accompanying DST.  An interesting, intrinsic feature of DST is that the macroscopic friction coefficient, $\mu$, {\it decreases} as the fraction of frictional contacts {\it increases}: the mean normal stress grows more rapidly than the shear stress.  This, and  contact network visualizations from simulations~\cite{mari2014shear}, indicates that there are important changes in the network of frictional contacts that are not captured by scalar variables such as  the fraction of frictional contacts.   In this work, we focus on the microscopic origin of the evolution of the components of the stress tensor across DST,  and construct a statistical theory for $\mu$, the anisotropy of the stress tensor.

 While the changes in real space near DST can be incremental,  and hence do not show any significant changes in pair correlations, the contact forces change dramatically and play a central role.
 The steady state flow of non-inertial suspensions is governed by microscopic constraints of force and torque balance, and these constraints can lead to non-trivial correlations of contact forces. Theories have focussed, up to now, on the {\it average} properties of the inter-particle forces \cite{wyart2014discontinuous}. However, fundamental questions about how interactions at the microscopic, contact level and the constraints of force balance give rise to a macroscopic transition remain \cite{sarkar2017shear}. 
 
In two-dimensional systems, the crucial constraint of force balance can be naturally accounted for by working in a {\it dual} space, known as a force tiling. In this representation, inter-particle forces are represented by the difference of vector {\it height fields}, $\lbrace \vec{h} \rbrace$, defined on the voids. This representation has been shown to be particularly useful in characterizing shear jamming transitions in frictional granular materials \cite{sarkar2013origin}.
Unlike shear jamming, where configurations and stresses are static,  flowing suspensions provide an {\it ensemble} of non-equilibrium steady states that are ripe for a statistical description. We show that the non-equilibrium steady states (NESS) at a given $\sigma_{xy}$ and $\phi$ can be mapped to  a {\it statistical  ensemble characterized by an a-priori probability distribution}.  This distribution is constructed from the measured pair correlation functions in force space.

In the continuum, the height fields define the local Cauchy stress tensor, by the relation $\tensor{\sigma} = \nabla \times \vec{h}$, and the area integral of  $\tensor{\sigma}$, or the force moment tensor, $\tensor{\Sigma}$~\cite{henkes2009statistical}, in terms of difference of the height fields across the system:
\begin{eqnarray}
\tensor{\sigma} = \left({\begin{array}{cc} \partial_y h_x & \partial_y h_y\\ -\partial_x h_x & -\partial_x h_y \end{array}}\right);~~
\tensor{\Sigma} = \left({\begin{array}{cc} L_y \Gamma_{yx} & L_y \Gamma_{yy}\\ -L_x \Gamma_{xx} & -L_x \Gamma_{xy} \end{array}}\right),
\label{stress_tensor_components_equation}
\end{eqnarray}
where $\vec{\Gamma}_{x(y)}$ represents the sum of forces along the $x(y)$ directions, and $L_{x(y)}$ represents the linear dimensions of the system  ($\tensor{\sigma} = \tensor{\Sigma}/L_xL_y$). Additionally, global torque balance implies $\Sigma_{xy}= \Sigma_{yx}$. In our simulations $L_x = L_y = L$, hence $\Gamma_{yy} = -\Gamma_{xx} = L \sigma_{xy} =  \sigma$.
Working with the ensemble of  force tilings generated from the NESS created in simulations, we observe changes in the anisotropy of the Pair Correlation Function of the Vertices (PCFV) of the tilings as $\phi$ and $\sigma_{xy}$ are changed. Using these microscopic correlations, we build a statistical theory for $\tensor{\Sigma}$. The reason for using the components of $\tensor{\Sigma}$  is  their clear geometric signatures in the force-tilings as shown in Fig. \ref{bodyforce_tile_figure}.
The stress anisotropy is defined as the ratio of the difference in eigenvalues, $\tau$, to the trace $2P = \sigma_{xx} +\sigma_{yy} $ of $\tensor{\sigma}$, which can also be related to the components of $\tensor{\Sigma}$:
\begin{equation}
\frac{\tau}{2P} = \frac{\sqrt{\tilde{N}_1^2 +4 \Sigma_{xy}^2 }}{\Sigma_{xx} + \Sigma_{yy}}~,
\label{mu_eqn}
\end{equation}
where $\tilde{N}_1 = \Sigma_{xx} - \Sigma_{yy}$. In the limit of $\tilde{N}_1 \rightarrow 0$, $\frac{\tau}{2P}$ is identical to the macroscopic friction coefficient $\mu = \frac{\sigma_{xy}}{P}$. In this letter, we show that the  change in the macroscopic friction coefficient, $\mu(\phi, \sigma_{xy})$, across the DST transition~\cite{SI} can be obtained from a statistical theory based on an effective pair potential between the vertices of the force tilings. An extension of the quasi-Newtonian, rate-independent, suspension rheology model~\cite{boyer2011unifying,PhysRevFluids.2.081301} can then be used to compute the viscosity, $\eta(\phi,\sigma_{xy})$:  \begin{equation}
\eta (\phi,\sigma_{xy}) \propto \mu(\phi,\sigma_{xy}) \left( \mu(\phi,\sigma_{xy})- \mu_c \right)^{-2} ~.
\label{eta_mu_relation}
\end{equation}
As we show~\cite{SI}, this constitutive relation is valid for thickening suspensions in the limit of $\phi \rightarrow \phi_m^-$, where $\phi_m$ is the frictional jamming point.  
We use our microscopic theory of $\mu$  in conjunction with this constitutive relation to predict the rheological properties characterizing  DST.
%
%
%


{\it Simulating Dense Suspensions:} 
We perform simulations of simple shear under constant stress of a 
monolayer of $N=2000$ bidisperse (radii $a$ and $1.4a$) 
spherical particles by methods described in detail previously~\cite{mari2014shear}.
These follow an overdamped dynamics and are subject to Stokes drag, 
pairwise lubrication, frictional contact, and short-range repulsive forces (see  Supplemental Information). 
Because of the repulsive force of maximum  $F_0$  at contact, frictional contacts only form 
for stresses about or larger than $\sigma_0 \equiv F_0/a^2$, which induces DST at volume fractions 
$\phi\gtrsim0.78$~\cite{mari2014shear}.

{\it Force Space Representation:}
For a force balanced configuration of grains with pairwise forces, the ``vector sum" of forces on every grain, i.e. the force vectors arranged head to tail (with a cyclic convention), form a {\it closed polygon}. Next, Newton's third law imposes the condition that every force vector in the system, has an equal and opposite counterpart that belongs to its neighboring grain. This leads to the force polygons being exactly edge-matching. Extending this to all particles within the system leads to a ``force tiling"~\cite{tighe2008entropy,sarkar2013origin}. The adjacency of the faces in the tiling is the adjacency of the grains, whereas the adjacency of the vertices is the adjacency of the voids (the heights
are associated with the voids in the network).
In addition to the pairwise forces between grains, each particle experiences a hydrodynamic drag, which can be represented as a {\it body force}. Imposing the constraints of vectorial force balance in the presence of body forces leads to a unique solution for modified height fields, given the geometrical properties of the contact network~\cite{ramola2017stress}.  This allows us to construct the ensemble of force tilings corresponding to the NESS of the suspension. The distribution of the hydrodynamic drag force to contacts through the modified height vectors leads to some very small contact forces, that do not represent ``real contacts''.  As we discuss below, we have a systematic way of neglecting these in our statistical analysis.
%

{\it Pair Correlation Functions:}
Using the force tiling representation, we compute the PCFV, defined to be
\begin{equation}
g_2(\vec{h}) = \left \langle \frac{A}{N_v(N_v-1)} \sum_{i = 1}^{N_v} \sum_{j \ne i}^{N_v} \delta\left(\vec{h} - (\vec{h}_i -\vec{h}_j)\right) \right \rangle,
\label{height_pair_correlation}
\end{equation}
where $N_v$ is the total number of voids in the system,  $A= |\vec{\Gamma}_x \times \vec{\Gamma}_y|$, and $\rho_v = N_v/A$ is the density of height vertices in the force tiling. The PCFV are averaged over $200$
configurations obtained from the simulated steady state
of dense suspensions at each $\phi$ and $\sigma_{xy}$~\cite{SI}.  We find a distinct fourfold anisotropic structure in $g_2(\vec{h})$, which quantitatively captures the details of the changes in the organization of the forces acting between particles as $\phi$ is increased (Fig. \ref{g2xy_figure}).  The anisotropy is sensitive, to a lesser extent, to  increases in $\sigma_{xy}$. The regions where $g_2(\vec{h}) < 1$ indicate regions of larger contact forces, statistically, since this is where the height vertices are farther apart than expected for an uncorrelated distribution. As seen from Fig. \ref{g2xy_figure}, these regions lie along the compressive direction for all values of $\phi$ and $\sigma_{xy}$.  Complementing these are the regions with $g_2(\vec{h}) > 1$,  which indicate regions of smaller forces.   The angles between these regions  clearly increase as $\phi$ increases~ \cite{SI}. 
These changes in $g_2(\vec{h})$, especially its anisotropy, have important consequences for the stress tensor, as we  show below.

{\it A Statistical Ensemble:}
Each force tiling is specified by a set of vertices and a
set of edges that connect these vertices. The distances
between the vertices quantify the internal stress in the
system, whereas the edges, which quantify the specific contact forces in a configuration, can be thought of, in a statistical sense, as fluctuating quantities, with connections between pairs of vertices  chosen with some weights. 
We thus treat these vertices of the force tilings as
the points of an interacting system of particles.   These effective interactions arise from the constraints of mechanical equilibrium, and from integrating out the edges.  We represent this effective interaction by a non-central potential computed from the measured pair correlation function, similar to constructions used in colloidal and polymer theory \cite{bolhuis2001accurate}:
\begin{figure}[t!]
\hspace*{-0.6cm}
\includegraphics[width=1.1\linewidth]{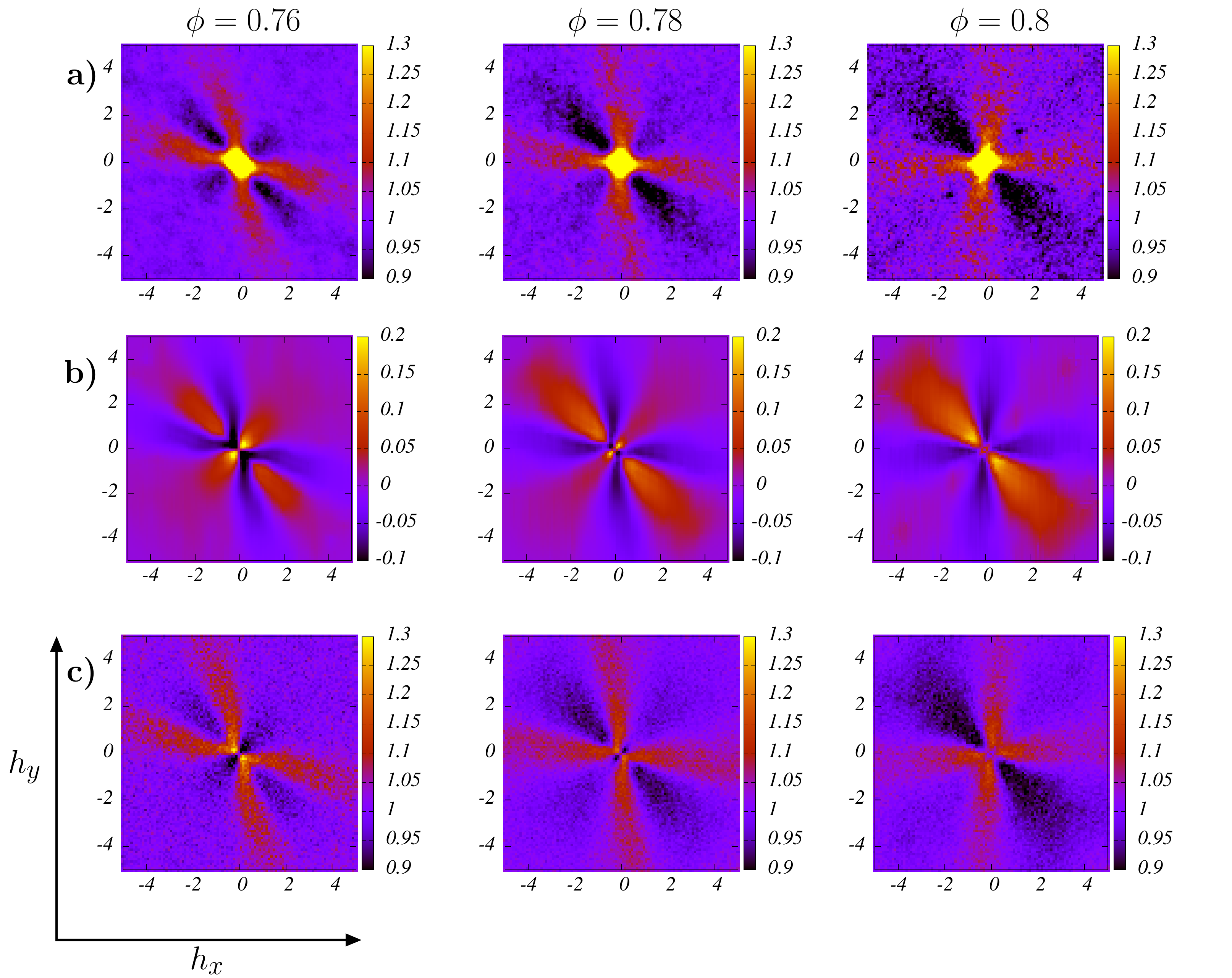}
\caption{(Color online) {\bf a)} Observed pair correlation functions at $\sigma_{xy} = 2\sigma_0$, at packing fractions $\phi = 0.76, 0.78$ and $0.8$. $\phi=0.8$ is above $\phi_{DST}$:  the onset packing fraction for a regime of stress over which the viscosity scales as $\sigma$, which defines DST (see \cite{SI}). The forces (and consequently the heights) have been scaled by the imposed shear stress. The change in symmetry of $g_2(\vec{h})$ is clearly visible as the packing fraction is increased. {\bf b)} Potentials constructed using these pair correlation functions (Eq. (\ref{potential_equation})). {\bf c)} A comparison with pair correlations obtained from direct Monte Carlo simulations of particles interacting via these potentials.
} 
\label{g2xy_figure}
\end{figure}
\begin{equation}
V_2(\vec{h}) = - \log \left( \frac{g_2(\vec{h})}{g_2(|\vec{h}|)}\right),
\label{potential_equation}
\end{equation}
The regularization through division by $g_2(|\vec{h}|)$ is necessary because there is strong clustering at very small distances in height space~\cite{SI}, which reflects the behavior of very small forces, much smaller than the repulsive force that needs to be overcome to create frictional contacts~\cite{mari2014shear,SI}. In addition, we add a short ranged repulsive potential to $V_2(\vec{h})$ that prevents clustering of vertices at the smallest force scales \cite{SI}. 
The resulting potential $V_{\phi,\sigma}(\vec{h})$ thus represents interactions at intermediate and large scales in the force tilings.   This potential encodes the full anisotropy of $g_2(\vec{h})$, and as we show below, this is crucial for understanding the evolution of the anisotropy of the stress tensor.   To check whether such a potential is successfully able to reproduce the original correlations, we perform Monte Carlo (MC) simulations, as described in detail in~\cite{SI}.  The $g_2(\vec{h})$ obtained from the MC simulations are shown in Fig. \ref{g2xy_figure}, and demonstrate that $V_2(\vec{h})$ captures the properties at all but the smallest force scales.

The force tiles obtained  from  the simulations form an ensemble with microstates defined by the set $\mathcal{C} \equiv \lbrace \vec h_i \rbrace$.  The fundamental assumption we make is that this ensemble of NESS is characterized by an {\it a priori} probability  $p(\mathcal{C}) \propto \exp(- V(\mathcal{C}))$, where $V(\mathcal{C}) = \sum_{i, j \ne i}$ $V_{\phi,\sigma}(\vec{h}_i - \vec{h}_j)$ is the analog of the total energy of a configuration in equilibrium statistical mechanics.  We then characterize the properties of the NESS by this generalized statistical ensemble.
The partition function of the system is then
\begin{eqnarray}
\nonumber
Z_{\phi, \sigma} = &&\frac{1}{N_v!}\int_{0}^{\infty} d A \exp \left( -N_v f_p^* A \right) \times \\
&&\underbrace{\int_{A} \prod_{i = 1}^{N_v} d \vec{h}_i \exp \left( - \sum_{i,j} V_{\phi, \sigma}(\vec{h}_i - \vec{h_j}) \right)}_{A^{N_v} \exp(-\epsilon_{\phi,\sigma}(A, N_v))}, \nonumber \\
&&=\int_{0}^{\infty} d A \exp (-\mathcal{F}_{A;\phi, \sigma}).
\label{partition_fn}
\end{eqnarray}
where the positions $\vec{h}_i$ are confined to be within the box with area $A$, which is related to stresses since this is the area of the {\it force tiling}.   
Here $f_p^*$ plays the role of a pressure in the ``NPT" ensemble in equilibrium statistical mechanics of particles, and controls the fluctuations of $A$.  Since $N_1$ is observed to be small in the simulations, we assume that it vanishes, which leads to the relationship $A = \sigma^2 \left( 1/\mu^2 - 1 \right)$ \cite{SI}.

We next construct a mean-field theory of $\mu$ by minimizing the effective ``free-energy'' function, $\mathcal{F}_{A;\phi, \sigma}$, referred to in the following as $\mathcal{F}$.
In order to compute $\mathcal{F}$, we sample $\epsilon_{\phi,\sigma}(A, N_v)$ (Eq. (\ref{partition_fn})). Details of the sampling method are provided in~\cite{SI}.
Transforming from $A$ to $\mu$,  the ``free energy" per vertex  is given by
\begin{eqnarray}
&&f(\mu;\phi, \sigma) \equiv \mathcal{F}/N_v \nonumber \\
&= & f_p^* \sigma^2\left(\frac{1}{\mu^2} \!\!- \!\!1\right)\!\! - \!\!\log \left[ \sigma^2\left(\frac{1}{\mu^2} \!\!- \!\!1\right) \right] \!\!+\!\!\frac{ \epsilon_{\phi,\sigma}\left( \mu,N_v \right)}{N_v}.~~~~
\label{free_energy_function_in_mu}
\end{eqnarray}
 As an example, the functions $f(\mu;\phi, \sigma)$ obtained at imposed stress $\sigma_{xy} = 100 \sigma_0$ at different packing fractions are shown in the inset of Fig. \ref{sampled_energy_figure}. We fix $f_p^* = 6.5 \times 10^{-4}$ to reproduce the observed value of $\mu$ at $\phi = 0.8$ and $\sigma_{xy} = 100\sigma_0$. 
\begin{figure}[t!]
\hspace*{-0.5cm}
\includegraphics[width=1.1\linewidth]{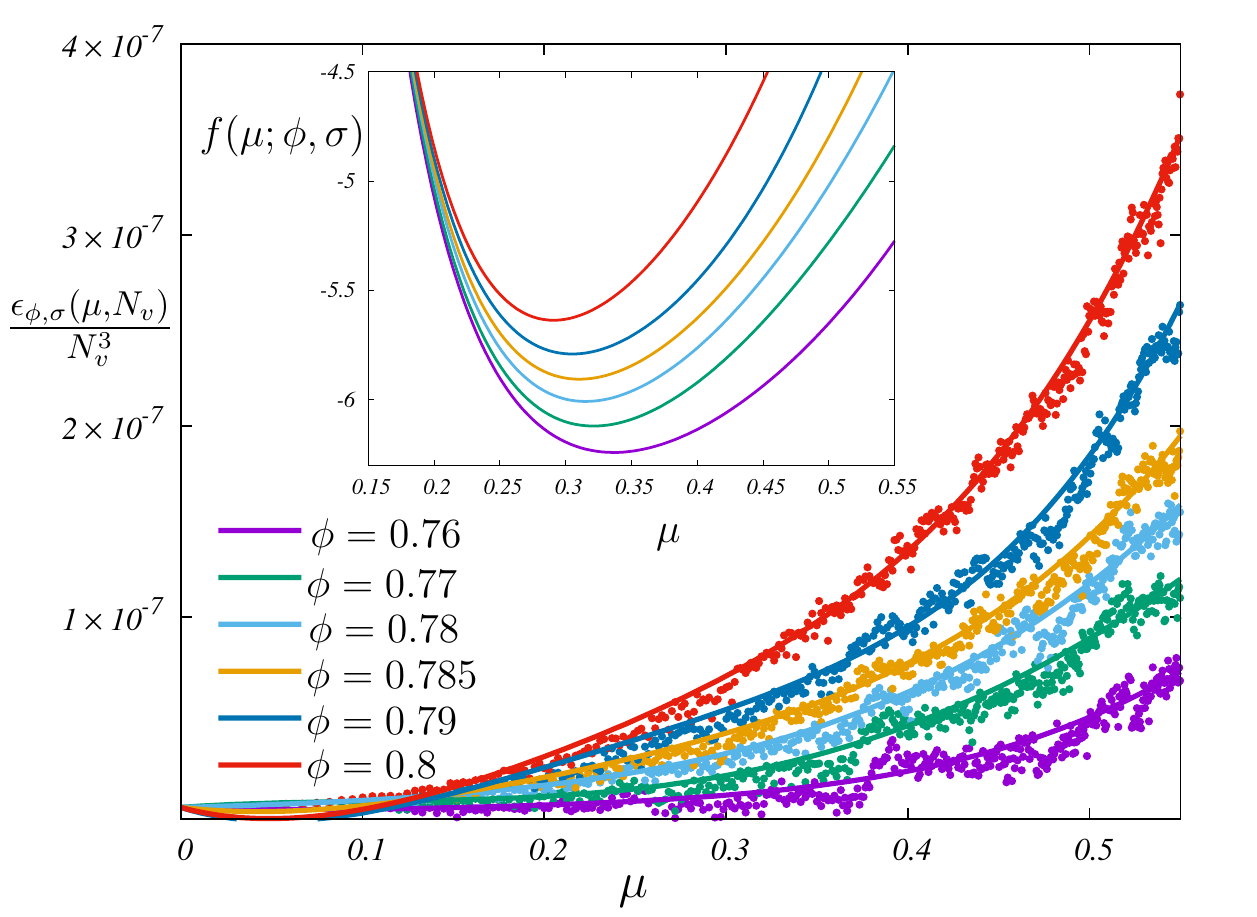}
\caption{(Color online) Sampled values of $\epsilon_{\phi,\sigma}(\mu, N_v)$ for $N_v = 1024$ and $\sigma_{xy}/\sigma_0 = 100$, with  $V_2$ derived from simulations at different packing fractions $\phi$ (Eq. (\ref{potential_equation})). (Inset) $f(\mu;\phi, \sigma)$  for  $N_v = 3000$, and $f_p^* = 6.5 \times 10^{-4}$. The minimum of $f(\mu;\phi, \sigma)$ provides the value of $\mu$ at each $(\phi, \sigma)$.} 
\label{sampled_energy_figure}
\end{figure}

{\it Phase Diagram for DST:}
Finally, minimizing $f(\mu;\phi, \sigma)$, we compute $\mu(\phi, \sigma)\equiv \mu(\phi, \sigma_{xy})$, and deduce the viscosity and the DST phase diagram.   
The variation of $\mu$ is provided in Fig. \ref{mu_variation_figure}. We find that $\mu$ decreases as the packing fraction $\phi$ and the confining shear stress $\sigma_{xy}$ are increased, in agreement with the variation observed directly in the simulations~\cite{SI}.  Unfortunately, there are no experimental measurements of $\mu(\phi, \sigma)$ in DST suspensions. However, insight may be gained from three-dimensional simulations of non-thickening suspensions where the second normal stress difference $N_2$ is found to be roughly linear with $P$ \cite{yurkovetsky2008particle}, and thus the behavior of $N_2$ gives a reasonable approximation of that of $P$. In particular, Cwalina and Wagner~\cite{cwalina2014material}  provide $N_2$ which is largely in agreement with the present simulation method~\cite{mari2015discontinuous}.
By the present simulation method applied to three-dimensional suspensions, $N_2/\sigma_{xy}$ increases (i.e. the ``friction coefficient" of $\sigma_{xy}/N_2$ decreases) at DST as seen in Fig. 6 of ref.~\cite{mari2014shear},  
  and thus it appears reasonable that the experimental ratio of $\sigma_{xy}/P$ also decreases at this transition.

The DST  boundary~\cite{SI} is defined by the condition $\frac{d\dot \gamma}{d \sigma_{xy}} = 0$.  This relationship, can be translated to one in terms of $\mu$ using Eq. (\ref{eta_mu_relation}):
\begin{equation}
\frac{\sigma_{xy}}{\mu}\Big|{ \frac{d \mu}{d \sigma_{xy}}} \Big|= \frac{\mu - \mu_c}{\mu + \mu_c}.
\label{mu_DST}
\end{equation}
Using the values of $\mu(\phi, \sigma)$ obtained by minimizing $f(\mu;\phi, \sigma)$,  we find that  Eq. (\ref{mu_DST}) is satisfied  at two values of the shear stress for $0.785 \le \phi \le 0.8$ if we choose $\mu_c$ to be $\mu(0.8,100)$ (Fig. \ref{mu_variation_figure}).  This choice implies that the viscosity diverges at $\phi = 0.8$ in the limit of large $\sigma$, where all contacts are frictional. The inset of Fig. \ref{mu_variation_figure} demarcates the DST region obtained from solving Eq. (\ref{mu_DST}). This region is not sensitive to the choice of $\mu_c$ as long as it is in the vicinity of the smallest value observed at $\phi \simeq 0.8$. The precise numerical values are not crucial as Eq. (\ref{mu_DST}) will have two solutions as long as the generic features in $g_2(\vec{h})$ that we obtain from the simulations are preserved. The results for $\eta$ as a function of $\phi$ and $\sigma_{xy}$ are shown in~\cite{SI}.

\begin{figure}[t!]
\includegraphics[width=1\linewidth]{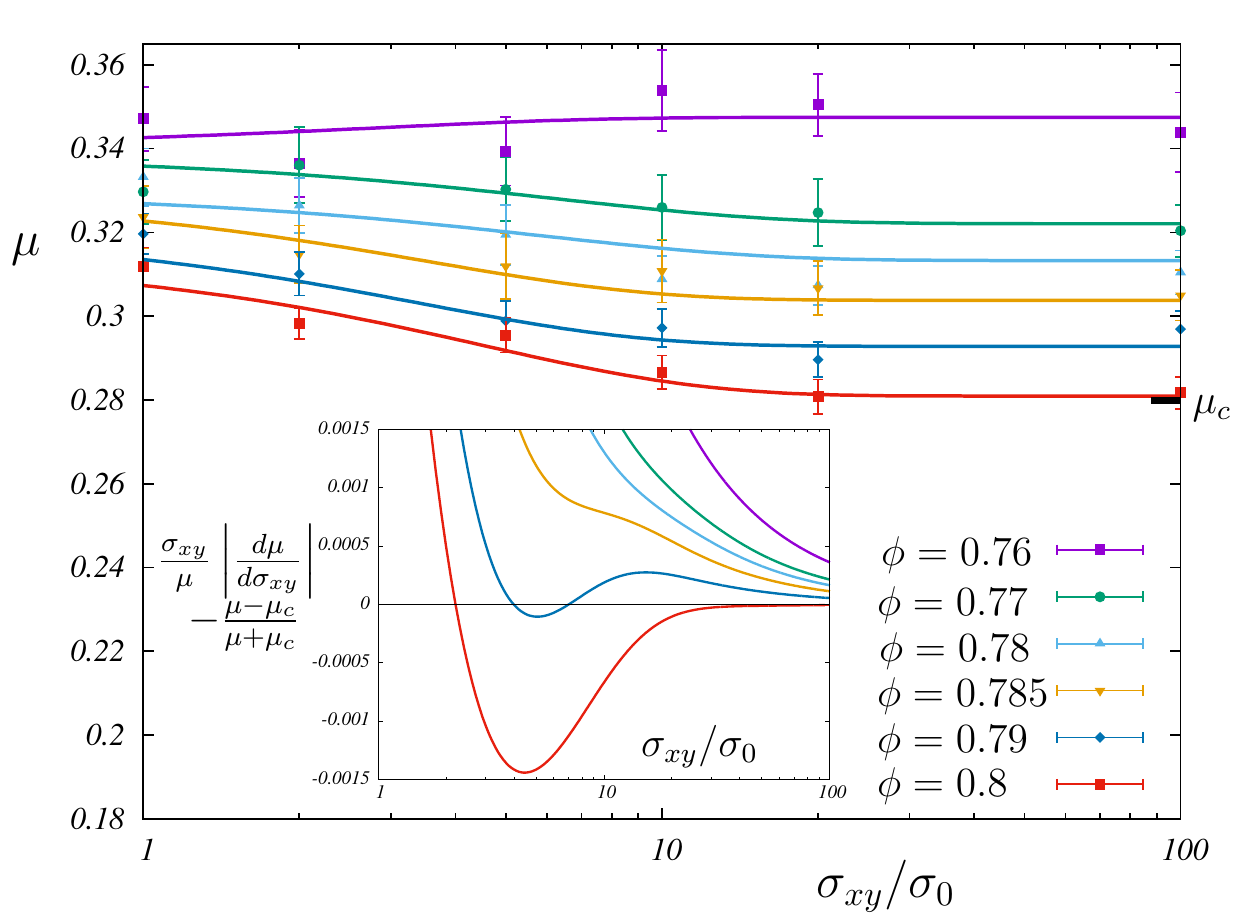}
\caption{(Color online) Variation of the macroscopic friction coefficient $\mu$, corresponding to the minimum of the free energy function in Eq. (\ref{free_energy_function_in_mu}). We find that $\mu$ decreases as packing fraction $\phi$ and the confining shear stress $\sigma_{xy}$ are increased.
(Inset) Plot of  Eq. (\ref{mu_DST}) showing the appearance of two solutions at $\phi=0.79$, and the second solution moving out to $\sigma_{xy} \rightarrow \infty$ at $\phi=0.8$.  }
\label{mu_variation_figure}
\end{figure}


{\it Conclusion and Outlook:} We have identified a correlation function that exhibits  significant changes in anisotropy across the DST transition. The correlations are in force space, and reflect the collective behavior triggered by changes in the nature of the {\it contact forces}, which often arise due to small changes in grain positions that are difficult to identify in any  positional correlations.  Remarkably, a theory based on pair potentials in force space describes the macroscopic rheology.  Our work also highlights the changes in the macroscopic friction coefficient, accompanying the DST transition.   The decrease in $\mu$ indicates that the pressure increase for an imposed increase of shear stress is larger in the  frictional branch of DST than  it is in the frictionless branch of DST~\cite{PhysRevFluids.2.081301}.   There is, however, no singular change in $\mu$ across the DST transition.  A decrease in $\mu(\phi, \sigma)$ has also been associated with the shear-jamming transition in dry grains~\cite{sarkar2016shear}. 
In that system,  overlap order parameters of the force tile vertices, evocative of spin glass order parameters,   characterized shear jamming~\cite{sarkar2016shear}.  In the DST steady states, these overlap parameters correspond to autocorrelation functions of the vertices of force tiles. In the future, we plan to use our statistical ensemble  to relate these autocorrelation functions to changes in viscosity accompanying  the DST transition.
Note that in equilibrium, stress autocorrelations  are related to the viscosity through the Green-Kubo relations.    

\vspace*{0.5cm}
{\it Acknowledements} The work of JT, KR, and BC has been supported by NSF-CBET-1605428, NSF-DMR-1409093 and the W. M. Keck Foundation. AS and JFM are supported under NSF-CBET-1605283. This research was also supported in part by the National Science Foundation under Grant No. NSF PHY11-25915.  We acknowledge the hospitality of the Kavli Institute for Theoretical Physics where part of this work was carried out.

\nocite{boyer2011unifying}
\nocite{wyart2014discontinuous,cates2014granulation}
\nocite{singh2017microstructural}
\nocite{mari2015nonmonotonic}
\nocite{mari2014shear}
\nocite{Laun_1994,mari2015discontinuous,mari2014shear}
\nocite{Cundall_1979,herrmann1998modeling}
\nocite{mari2014shear,Singh_2015}
\nocite{wyart2014discontinuous,singh2018constitutive}
\nocite{panagiotopoulos1987direct}
\nocite{yashonath1985monte}

\bibliography{bib_DST} 
\bibliographystyle{apsrev4-1}

\clearpage

\begin{widetext}

\begin{appendix}
 

\section*{\large Supplemental Material for ``Microscopic origin of frictional rheology in dense suspensions:  correlations in force space"}

In this document we provide supplemental figures and details of the calculations presented in the main text.

\maketitle


\subsection{ Macroscopic Friction Coefficient and DST Rheology}
The  existing mean-field  theory of DST extends the suspension rheology framework ~\cite{boyer2011unifying} through the introduction of a stress- and $\phi$-dependent microstructure parameter: the fraction of frictional contacts~\cite{wyart2014discontinuous,cates2014granulation}. The suspension rheology model embodies a constitutive relation:  $\mu(\phi,I_v)$, where the viscous number $I_v \equiv  \frac{\eta_f \dot \gamma}{P}$.   In this framework, the shear viscosity of the suspension \cite{boyer2011unifying} is: $\eta = \frac{\mu(I_v(\phi))}{I_v(\phi)}$, and $I_v (\phi) \propto (\phi_m -\phi)^2$, where $\phi_m$ is the jamming packing fraction at which $\eta$ diverges.  In the jamming limit,  $I_v\rightarrow 0$, one can also write a relationship between $\mu$ and $I_v$ (Eq. 5 in Ref.~\cite{boyer2011unifying}): $\mu - \mu_c \simeq I_v^{1/2}$, where, $\mu_c$ is a material parameter \cite{boyer2011unifying}.  Using this, an equivalent expression for the viscosity is: $\eta \propto \mu(\mu-\mu_c)^{-2}$, which focuses on the divergence of the viscosity of frictional suspensions as $\mu \rightarrow \mu_c^+$.  This is a consistent picture of the rate-independent, quasi-Newtonian rheology for a given microscopic friction coefficient.
\begin{figure}[h!]
\includegraphics[width=0.5\linewidth]{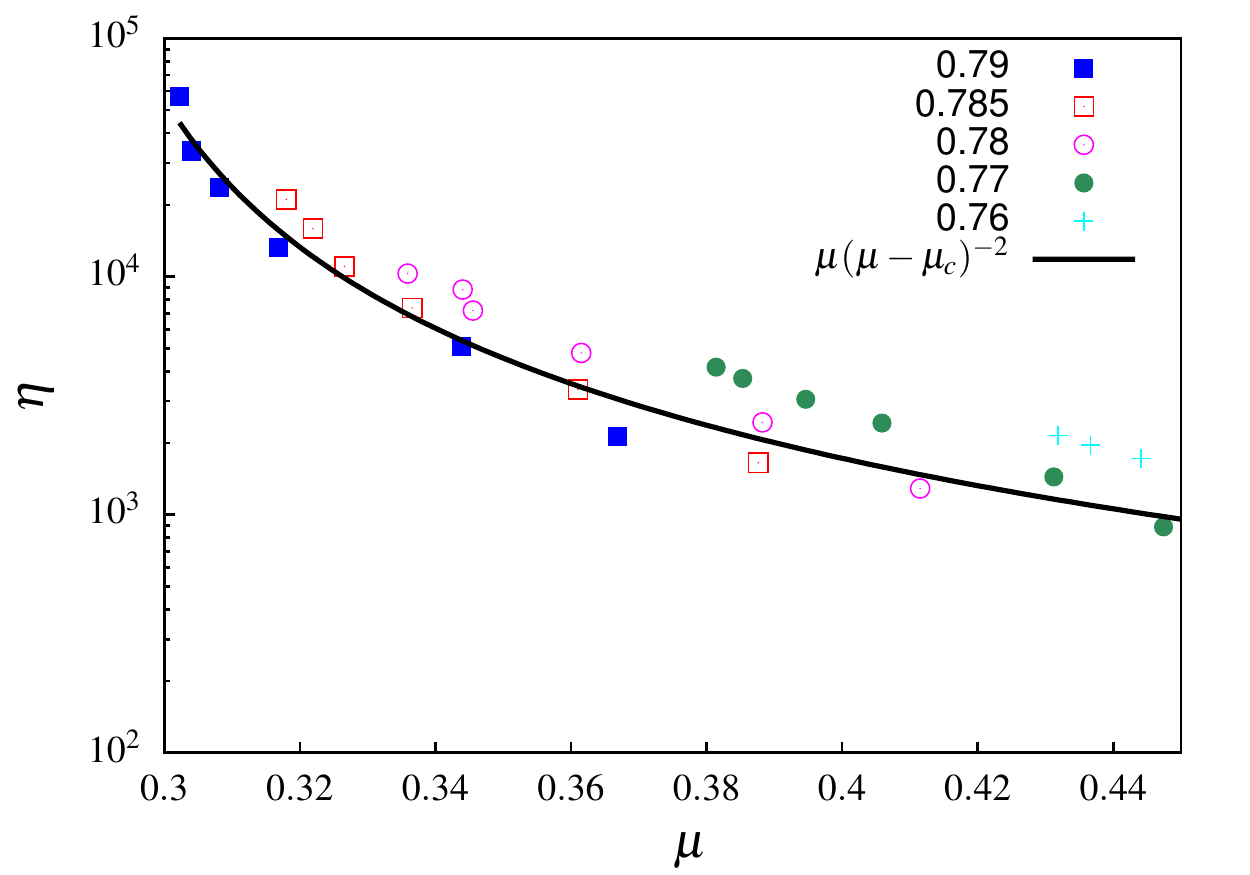}

\caption{ (Color online)  Plot of the viscosity, $\eta (\phi,\sigma_{xy})$ vs $\mu(\phi,\sigma_{xy})$ for different packing fractions, obtained from the simulations (symbols) compared to the constitutive relation: Eq. (\ref{eta_mu_SI}). Here  $\mu_c = 0.285$, is chosen to be the lowest value of the stress anisotropy observed in the simulations. The viscosity $\eta$ is measured in units of $\eta_0$, the viscosity of the underlying Newtonian  fluid, and in our simulations we set $\eta_0 = 1$.
} 
\label{flowcurves}
\end{figure}

Below, we extend this theory of rate-independent, quasi-Newtonian rheology to dense suspensions.   The physical picture underpinning the theory is the same as the mean-field theory of DST~\cite{wyart2014discontinuous,cates2014granulation}: frictional contacts increase with increasing imposed shear stress.  In our theory, the effect of this increase is represented by the ``order parameter'' $\mu(\phi,\sigma_{xy})$.   The theory for this order parameter is based on an effective pair potential in force space, as described in the main text.  We propose that the viscosity has the same functional dependence on $\mu$ as in the rate-independent suspension rheology but the physics of  thickening suspensions is encapsulated in the order parameter, $\mu(\phi,\sigma_{xy})$.  The viscosity of a thickening suspension  should diverge as $\phi \rightarrow \phi_m^-$, the jamming packing fraction  of the frictional fluid~\cite{wyart2014discontinuous,cates2014granulation,boyer2011unifying},  in the limit of $\sigma_{xy} \rightarrow \infty$ where the fraction of frictional contacts approaches unity.  Therefore, we define $\mu_c = \mu(\phi=0.80, \sigma_{xy} = 100 \sigma_0)$, the value we obtain from the theory at the highest packing fraction and shear stress.   Thus:
\begin{equation}
\eta (\phi,\sigma_{xy}) \propto \mu(\phi,\sigma_{xy}) \left( \mu(\phi,\sigma_{xy})- \mu_c \right)^{-2} ~.
\label{eta_mu_SI}
\end{equation}
The above constitutive relation is expected to be valid only close to $\mu_c$, and as it is approached from above.   In Fig. \ref{flowcurves}, we show that the increase in  $\eta (\phi,\sigma_{xy})$ is  primarily controlled by the decrease in $\mu(\phi,\sigma_{xy})$, close to $\mu_c$.   The functional form given in Eq. (\ref{eta_mu_SI}) is also seen to provide a good description of this correlation for the larger values of $\phi$.  We, therefore, use Eq. (\ref{eta_mu_SI}) to infer the rheological properties and compute the DST diagram.
The difference with the Wyart-Cates theory is that we encapsulate the information about the microstructure in $\mu(\phi,\sigma_{xy})$ rather than in the fraction of frictional contacts~\cite{wyart2014discontinuous,cates2014granulation}.

\section*{Simulating Dense Suspensions}
We simulate a two-dimensional or monolayer suspension of non-Brownian spherical particles
immersed in a Newtonian fluid under an imposed shear stress $\sigma_{xy}$.
This gives rise to a velocity field $\vec{v} = \dot\gamma(t)\hat{\vec{v}}(\vec{x}) = \dot\gamma(t)(x_2, 0)$
\cite{singh2017microstructural}, with a time-dependent shear rate $\dot\gamma$~\citep{mari2015nonmonotonic}.
All our results are obtained with $N=2000$ particles in a unit cell with Lees-Edwards boundary conditions.
Bidispersity at a radii ratio of $a$ and $1.4a$ and volume ratio of $1:1$ is used to
avoid crystallization during flow \cite{mari2014shear}.
In this simulation scheme, the particles interact through near-field hydrodynamic interactions (lubrication), a short-ranged repulsive force
and frictional contact forces. 

The motion is considered to be inertialess, so that the equation of motion reduces to a force balance between
 hydrodynamic ($\vec{F}_{\mathrm{H}}$), repulsive ($\vec{F}_{\mathrm{R}}$), and contact ($\vec{F}_{\mathrm{C}}$) forces:
 \begin{equation}
  0 = \vec{F}_{\mathrm{H}}(\vec{X},\vec{U}) + \vec{F}_{\mathrm{C}}(\vec{X}) + \vec{F}_{\mathrm{R}}(\vec{X}), 
  \label{eq:force_balance}
\end{equation}
where $\vec{X}$ and $\vec{U}$ denote particle positions and their velocities/angular velocities respectively.

The translational velocities and rotation rates are made dimensionless with $\dot{\gamma}a$ and $\dot{\gamma}$, respectively. 
The hydrodynamic forces are the sum of a drag due to the motion
relative to the surrounding fluid and a resistance to the deformation
imposed by the flow:
\begin{equation}
  \vec{F}_{\mathrm{H}}(\vec{X},\vec{U}) =
  -\tensor{R}_{\mathrm{FU}}(\vec{X}) \cdot \bigl(\vec{U}-\dot\gamma\hat{\vec{U}}^{\infty} \bigr)
  + \dot\gamma\tensor{R}_{\mathrm{FE}}(\vec{X}):\hat{\tensor{E}}, \label{eq:hydro_force}
\end{equation}
with $\hat{\vec{U}}^{\infty} = (\hat{\vec{v}}(y_1), \dots, \hat{\vec{v}}(y_N), \hat{\vec{\omega}}(y_1), \dots, \hat{\vec{\omega}}(y_N))$
and $\hat{\tensor{E}} = (\hat{\tensor{e}}(y_1), \dots, \hat{\tensor{e}}(y_N))$.

Details about the position-dependent resistance tensors
$\tensor{R}_{\mathrm{FU}}$ and $\tensor{R}_{\mathrm{FE}}$ are available in~\citep{mari2014shear}.
We regularize the resistance matrix by introducing a small cutoff length scale $\delta=10^{-3}$ \cite{mari2014shear}.
%

\begin{figure}[ht!]
\includegraphics[width=0.6\linewidth]{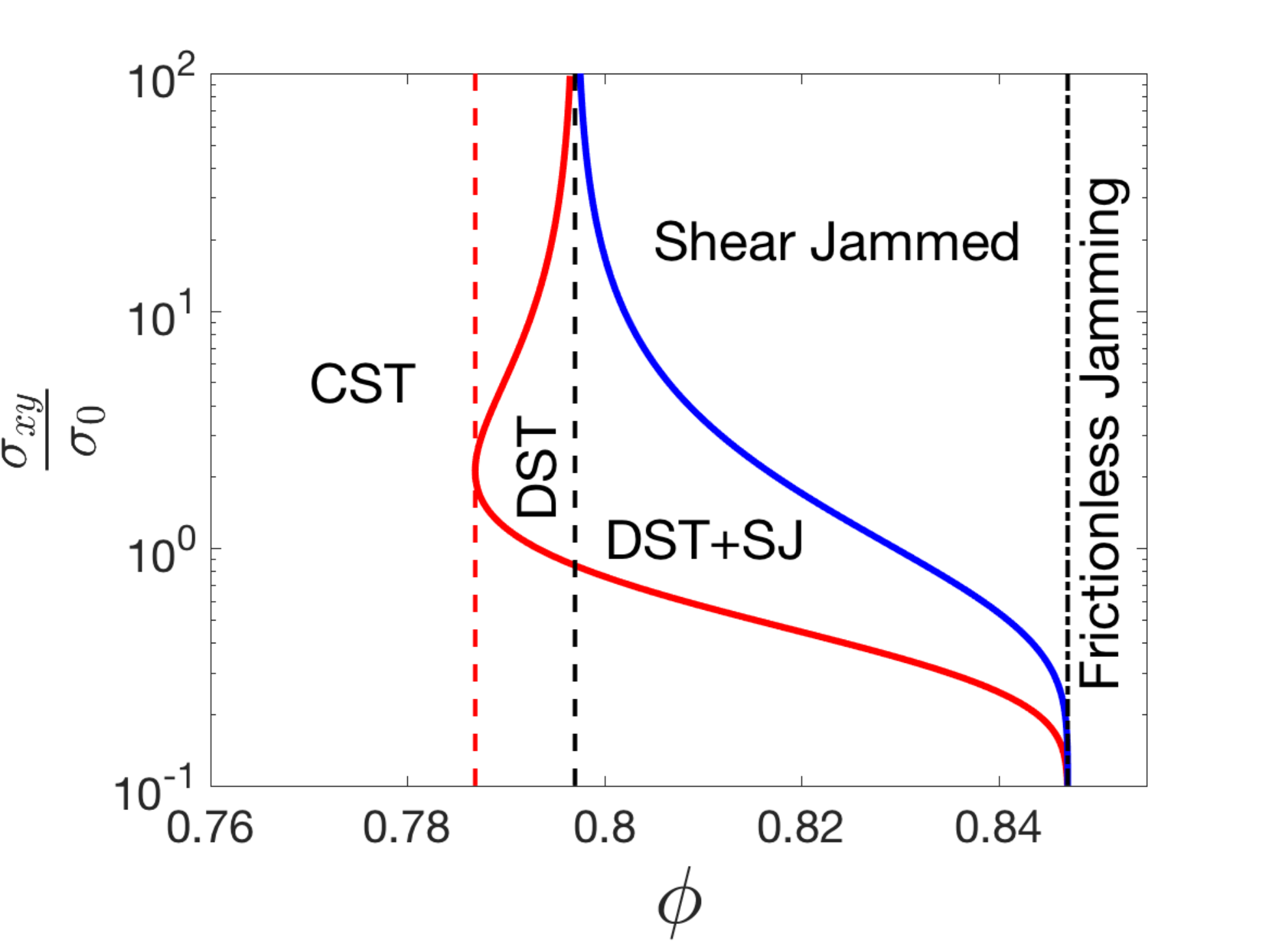}
\caption{(Color online)
Phase diagram in the shear stress--packing fraction $(\sigma_{xy},\phi)$ plane. The left (red)
curve locates the points where $\frac{d{\dot{\gamma}}}{d\sigma_{xy}} = 0$. The right (blue)
curve shows packing fraction dependent maximal stress above which the suspension is shear-jammed,
i.e., above which no flowing states exist. Dashed and dotted-dashed black lines represent frictional
and frictionless jamming points, respectively. 
The red dashed line shows the minimum packing fraction $\phi_{\rm DST}$ at which DST is observed. The regime of stress over which the viscosity scales as $\sigma$, defines the DST region.
} 
\label{phase_diagram}
\end{figure}


We take a stablizing repulsive force which decays exponentially with the
interparticle gap $h$ as $|\vec{F}_R|  = F_0 \exp(-h/\lambda)$,
with a characteristic length $ \lambda$.
This provides a simple model of screened electrostatic
interactions which can often be found in aqueous systems~\citep{Laun_1994,mari2015discontinuous,mari2014shear},
in which case $\lambda$ is the Debye length.
%
%
In the simulations, we set $\lambda = 0.02a$.
%


We model contact forces using linear springs and dashpots, a model that is commonly used in soft-sphere DEM simulations
~\citep{Cundall_1979,herrmann1998modeling}; the spring constants used here have a ratio $k_{\rm t} = 0.5 k_{\rm n}$.
For each applied stress, 
we adjust the spring stiffnesses such that the maximum particle overlaps do not
exceed 3{\%} of the particle radius in order to stay close to the rigid limit \cite{mari2014shear,Singh_2015}.
The normal and tangential components of the contact force
$\vec{F}_C^{(ij)}$ fulfill Coulomb's friction
law  $|F_{C,t}^{(ij)}| \le \mu_f |F_{C,n}^{(ij)}|$, where $\mu_f$ is the interparticle friction coefficient.
In this study we use $\mu_f=1.0$.

The unit scales for strain rate and stress are $\dot{\gamma}_0 \equiv F_{\rm 0}/{6\pi \eta_0 a^2}$
and $\sigma_0 \equiv \eta_0 \dot{\gamma}_0 = \frac{F_{\rm 0}}{{6\pi a^2}}$, respectively, where $\eta_0$ is the viscosity of the underlying Newtonian  fluid, and in our simulations we set $\eta_0 = 1$.

Based on the simulation results presented here and the model proposed in~\citep{wyart2014discontinuous,singh2018constitutive},
a phase diagram in $(\sigma,\phi)$ plane is displayed in Fig.~\ref{phase_diagram}.
For low packing fraction $\phi < \phi_{\rm DST}$, CST is observed.
For packing fractions, $ \phi_{\rm DST} \le \phi < \phi_{\rm J}^\mu$, DST is observed between two flowing states. In this range of $\phi$,
red curve shows locus of DST points, i.e., $\frac{d{\dot{\gamma}}}{d\sigma_{xy}} = 0$. For $\phi > \phi_{\rm J}^\mu$, DST is
 observed between a flowing and solid--like shear jammed state. The stress required to observe DST as well as shear jamming decreases
 with increase in packing fraction and both eventually vanish on the approach to the isotropic jamming  point.


\subsection{Dimensions of the Force Tiling Box}
Keeping the shear stress, $\sigma_{xy} = \sigma_{yx}$ and the real space dimensions, $L_x=L_y = L$ fixed implies that we fix
\begin{equation}
 \Gamma _{\text{yy}} = -\Gamma_{\text{xx}} = \sigma~.
 \label{sigma_definition_SI}
\end{equation}
We define
\begin{equation}
\mathcal{N}_1 =   \Gamma_{\text{yx}}+\Gamma _{\text{xy}},
\label{N1_force_tile}
\end{equation}
and
\begin{equation}
\mathcal{P} =   \Gamma_{\text{yx}}-\Gamma _{\text{xy}}.
\label{P_force_tile}
\end{equation}
The behaviour of these two quantities as $\phi$ and $\sigma_{xy}$ are varied are shown in Figs. \ref{pressure_figure} and \ref{n1_figure}. In addition, we plot the density of vertices $\rho_v = N_v/A$, where $N_v$ is the number of vertices, and $A = \left| \vec{\Gamma}_x \times \vec{\Gamma}_y \right|$ is the area of the force tiling box, as $\phi$ and $\sigma_{xy}$ are varied in Fig. \ref{density_figure}.

\begin{figure}[h!]
\hspace*{-1.2cm}
\includegraphics[width=0.5\linewidth]{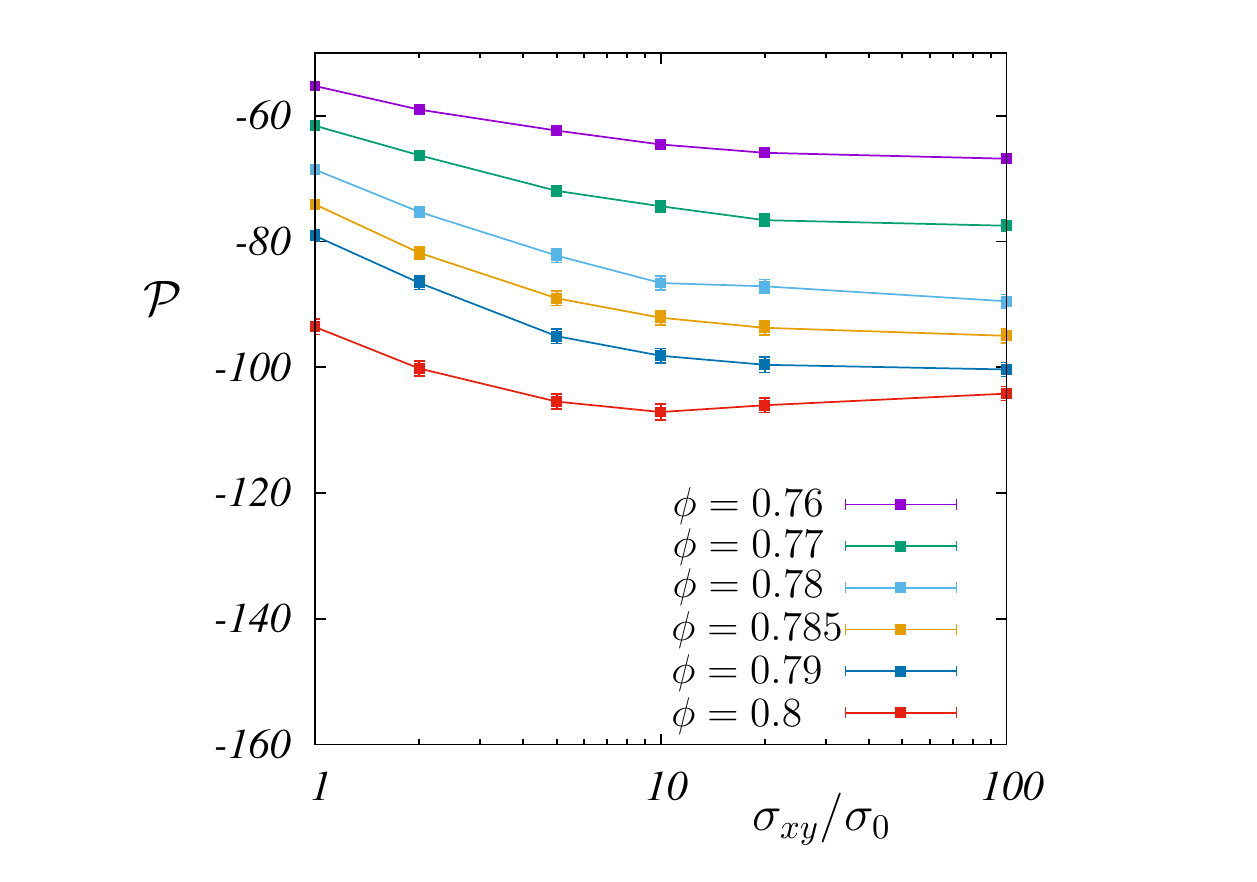}
\hspace*{-1.2cm}
\includegraphics[width=0.5\linewidth]{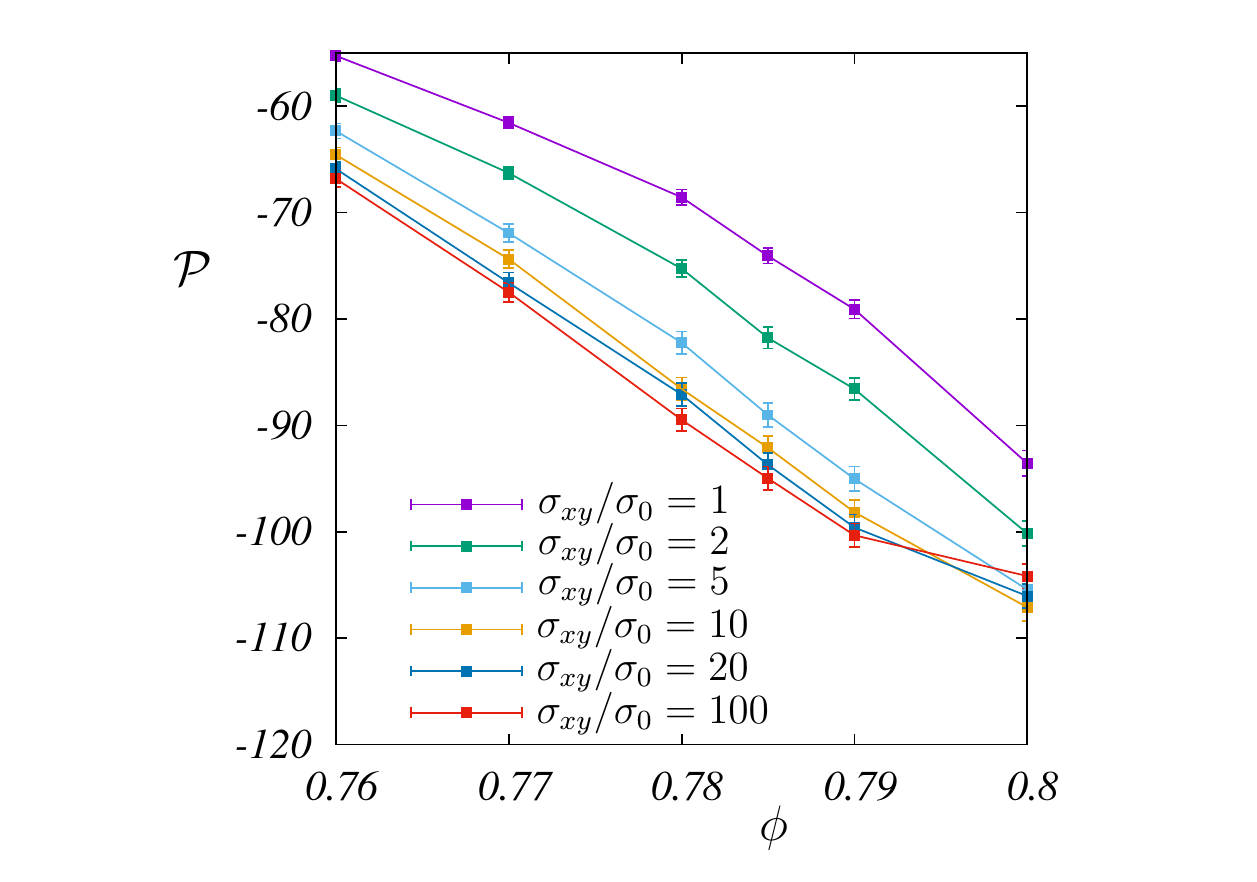}

\caption{ (Color online) Observed $\mathcal{P}$ from the data.} 
\label{pressure_figure}
\end{figure}

\begin{figure}[h!]
\hspace*{-1.2cm}
\includegraphics[width=0.5\linewidth]{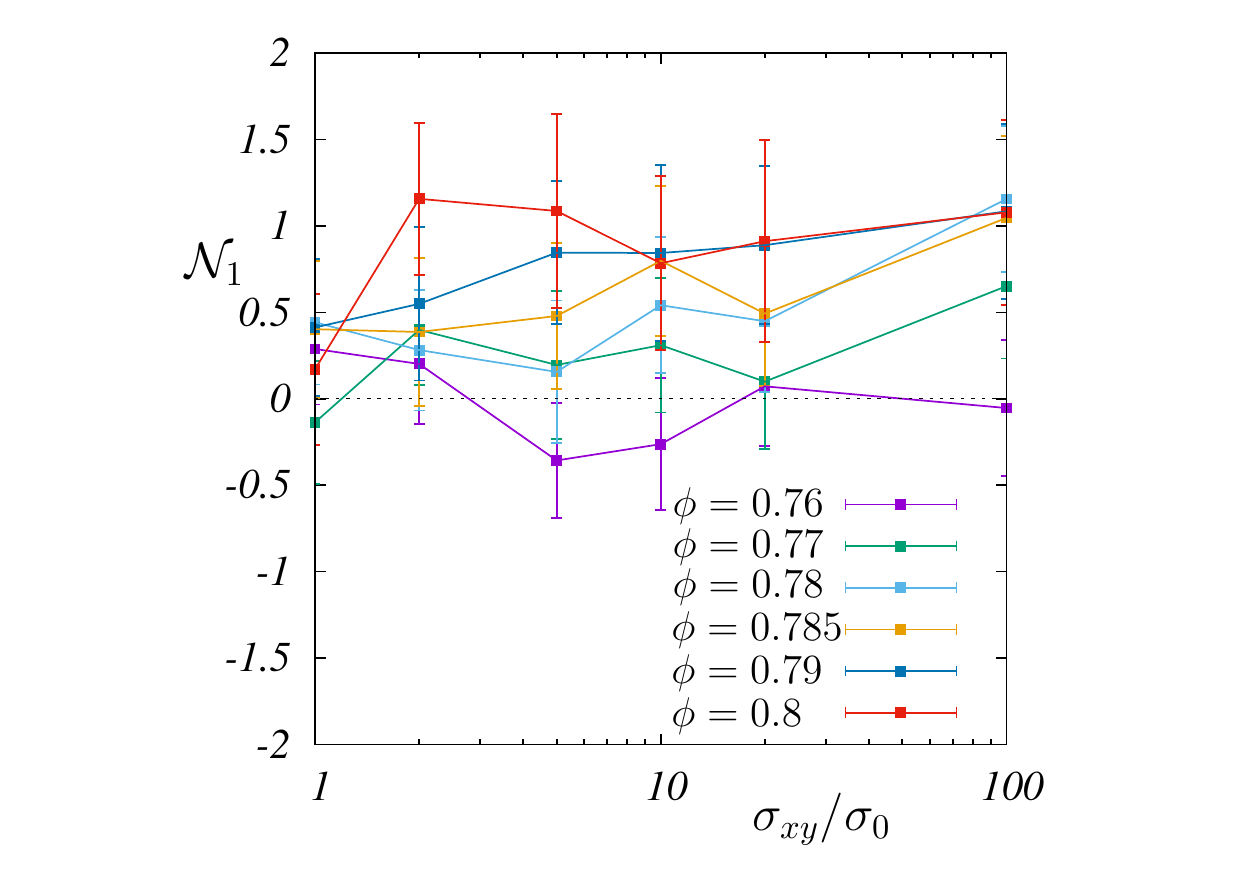}
\hspace*{-1.2cm}
\includegraphics[width=0.5\linewidth]{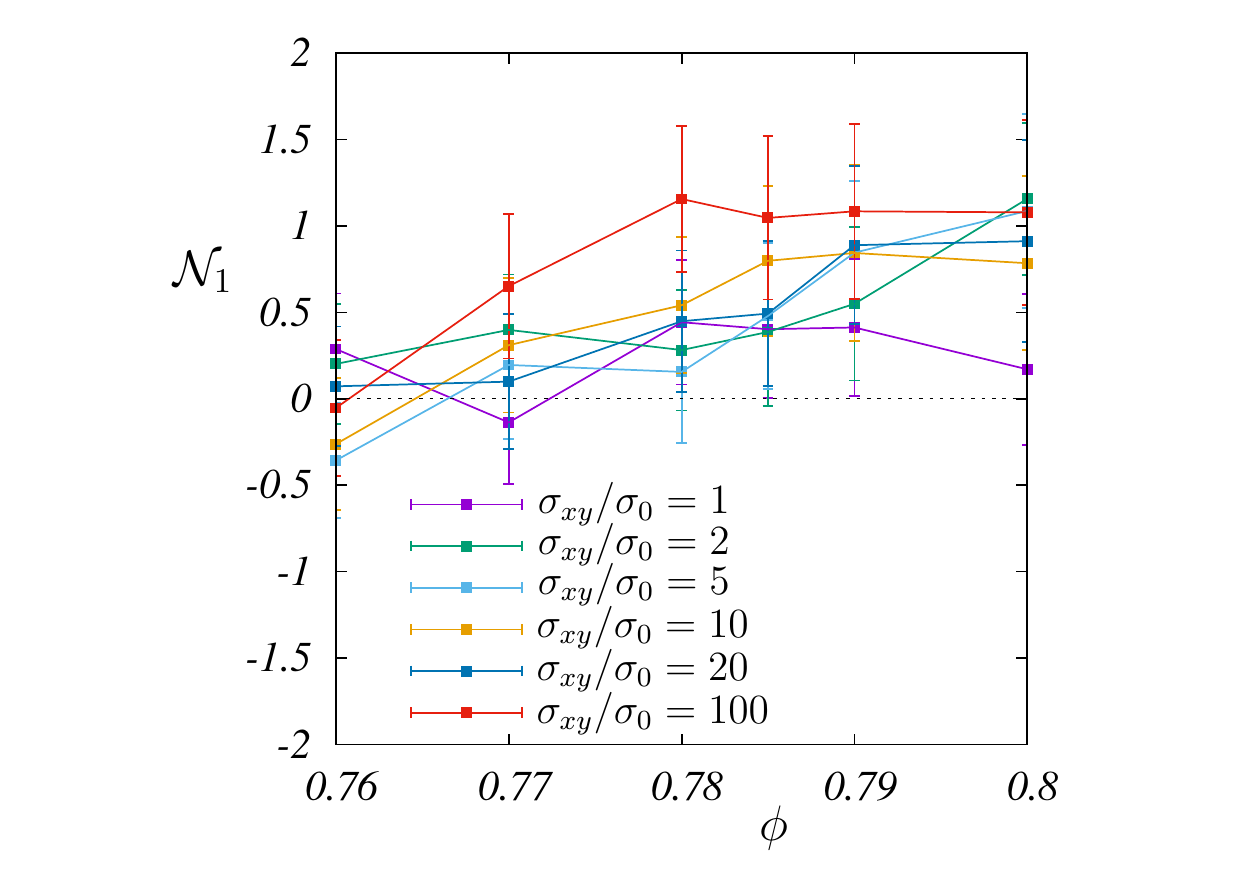}

\caption{ (Color online) Observed $\mathcal{N}_1$ from the data.
} 
\label{n1_figure}
\end{figure}

\begin{figure}[h!]
\hspace*{-1.2cm}
\includegraphics[width=0.5\linewidth]{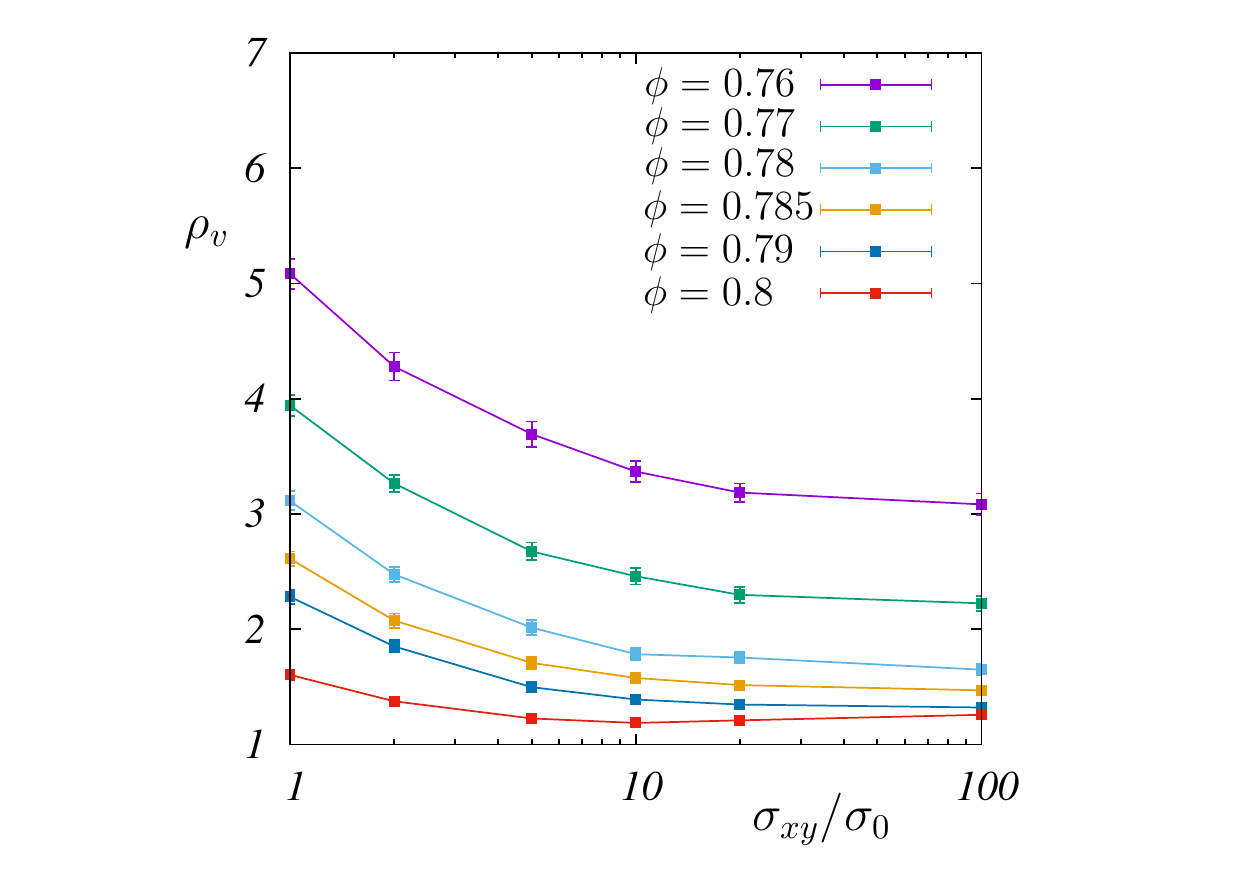}
\hspace*{-1.2cm}
\includegraphics[width=0.5\linewidth]{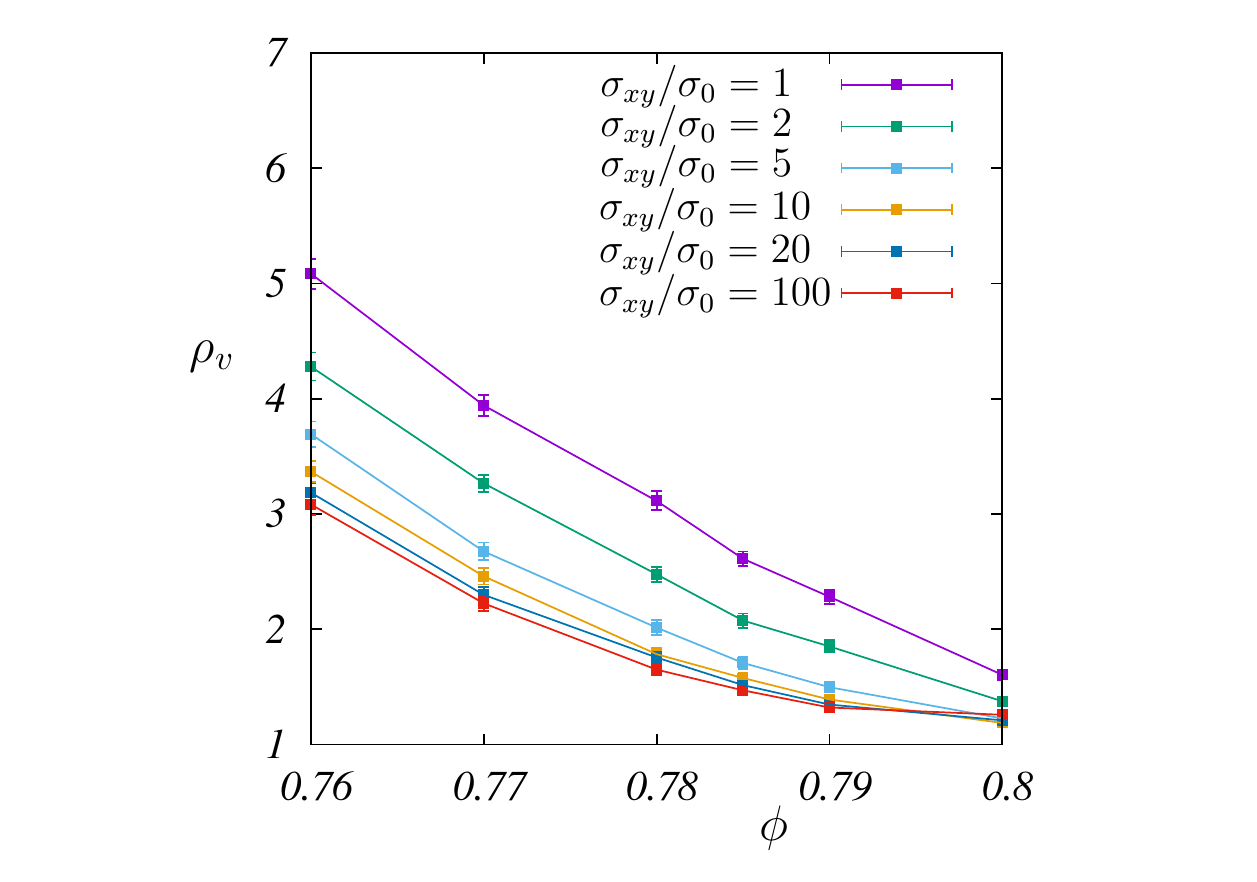}

\caption{ (Color online) Observed density of vertices in the force tiling, $\rho_v = N_v/A$ from the data.}
\label{density_figure}
\end{figure}

\vspace{2cm}
\subsection{Constraints on the Stress and Force Moment Tensors}
From Eq. (1) in the main text, the stress tensor $\tensor{\sigma}$ is given by
\begin{eqnarray}
\tensor{\sigma} = \left({\begin{array}{cc} \sigma_{xx} & \sigma_{xy}\\ \sigma_{yx} & \sigma_{yy} \end{array}}\right) = \frac{1}{L^2}\tensor{\Sigma} = \frac{1}{L^2}\left({\begin{array}{cc} \Sigma_{xx} & \Sigma_{xy}\\ \Sigma_{yx} & \Sigma_{yy} \end{array}}\right)  = \frac{1}{L}\left({\begin{array}{cc} \Gamma_{\text{yx}} &  \Gamma_{\text{yy}}\\ - \Gamma_{\text{xx}} & - \Gamma_{\text{xy}} \end{array}}\right).
\label{stress_tensor_equation_SI}
\end{eqnarray}
Global torque balance implies $\tensor{\sigma}^T = \tensor{\sigma}$, hence 
$\sigma_{xy} = \sigma_{yx}$.
The eigenvalues of $\tensor{\sigma}$ are then given by
\begin{eqnarray}
\lambda_{\pm}
&=&\frac{1}{2} \left( \left(\sigma_{xx} + \sigma_{yy} \right) \pm\sqrt{\left(\sigma_{xx} - \sigma_{yy} \right)^2 +4 \sigma_{xy}^2 }\right)\\
&=& \frac{1}{2 L^2} \left( \left(\Sigma_{xx} + \Sigma_{yy} \right) \pm\sqrt{\left(\Sigma_{xx} - \Sigma_{yy} \right)^2 +4 \Sigma_{xy}^2 }\right).
\end{eqnarray}
The normal stress difference is given by
\begin{equation}
N_1 =\sigma_{xx} - \sigma_{yy} = \frac{1}{L^2}(\Sigma_{xx} - \Sigma_{yy}) = \frac{1}{L^2}\tilde{N}_1.
\end{equation}
Using Eq. (\ref{N1_force_tile}) we have
\begin{equation}
\tilde{N}_1 = L  (\Gamma _{\text{yx}}+ \Gamma _{\text{xy}}) = L \mathcal{N}_1.
\end{equation}
The difference in the eigenvalues of the stress tensor is given by 
\begin{equation}
\tau = \frac{1}{L^2}\sqrt{(\tilde{N}_1)^2 +4 \Sigma_{xy}^2 } = \frac{1}{L}\sqrt{(\mathcal{N}_1)^2 +4 \sigma^2},
\end{equation}
where we have used Eqs. (\ref{sigma_definition_SI}) and (\ref{N1_force_tile}) in the last equality.
The sum of the eigenvalues is given by
\begin{equation}
2 P = \sigma_{xx} + \sigma_{yy} = \frac{1}{L^2} (\Sigma_{xx} + \Sigma_{yy}) = \frac{\mathcal{P}}{L},
\end{equation}
where $P$ is the pressure, and we have used Eq. (\ref{P_force_tile})  in the last equality.
The stress anisotropy, defined as the ratio of the difference of the eigenvalues ($\tau$) to the sum of the eigenvalues ($2P$) of the stress tensor can then be expressed as 
\begin{equation}
\frac{\tau}{2P} = \frac{\sqrt{\left(\tilde{N}_1\right)^2 +4 \Sigma_{xy}^2 }}{\Sigma_{xx} + \Sigma_{yy}} = \frac{\sqrt{(\mathcal{N}_1)^2 +4 \sigma^2}}{\mathcal{P}},
\end{equation}
which is Eq. (2) in the main text.
Since $\mathcal{N}_1/\mathcal{P}$ is observed to be small (Figs. \ref{pressure_figure} and \ref{n1_figure}), the stress anisotropy is 
\begin{equation}
\frac{\tau}{2P} \approx \frac{2 \sigma} {\mathcal{P}} = \frac{\sigma_{xy}}{P} = \mu.
\end{equation} 
The behaviour of $\mu$ observed from the simulations as $\phi$ and $\sigma_{xy}$ are varied is shown in Fig. \ref{mu_figure}.
Finally, if we set $\mathcal{N}_1 =0$, the area of the bounding box of the force tiles is given by
\begin{equation}
A=  \left| \vec{\Gamma}_x \times \vec{\Gamma}_y \right| = \Gamma _{\text{xx}} \Gamma _{\text{yy}}-\Gamma _{\text{xy}} \Gamma _{\text{yx}} = \sigma^2 \left(\frac{1}{\mu^2} - 1 \right).
\end{equation}
\vspace*{-0.5cm}
\begin{figure}[h!]
\hspace*{-1.2cm}
\includegraphics[width=0.5\linewidth]{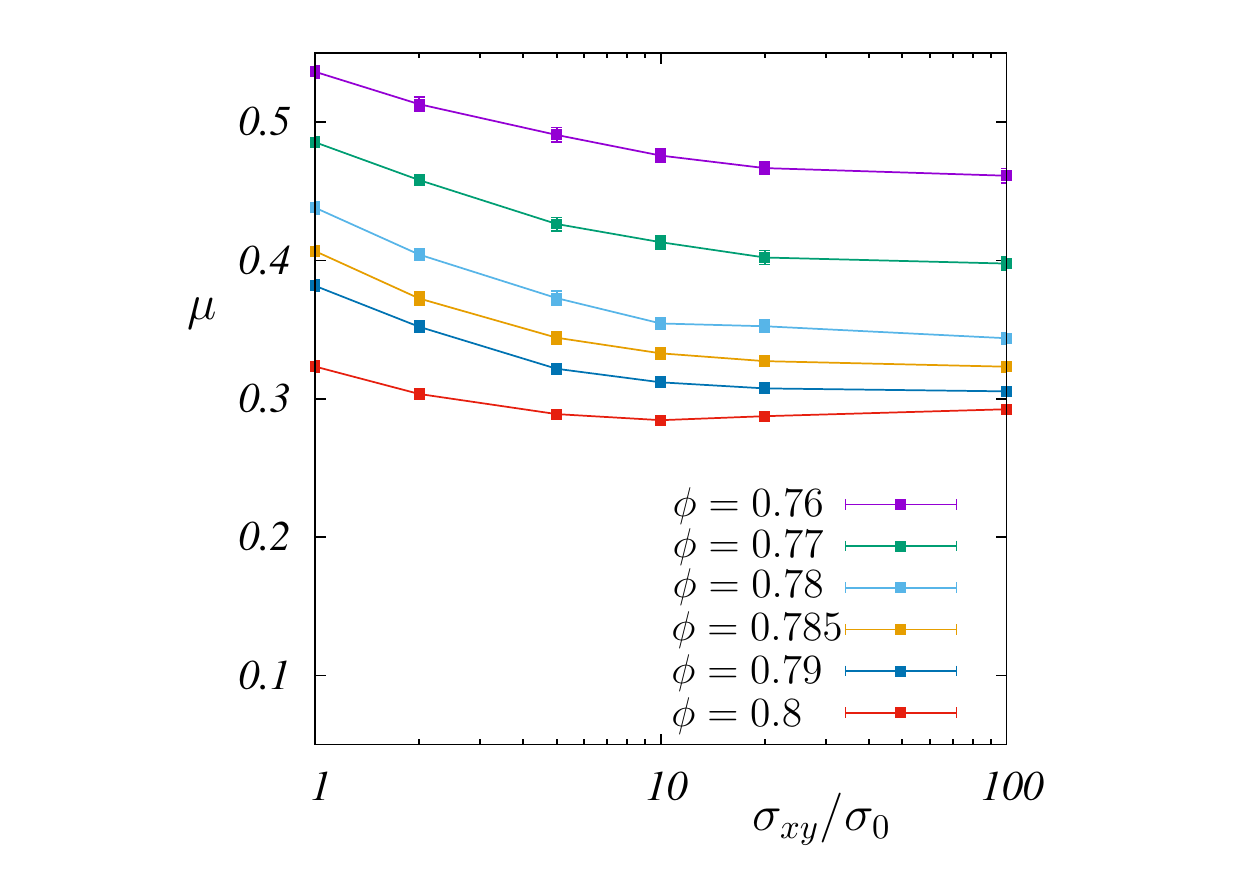}
\hspace*{-1.2cm}
\includegraphics[width=0.5\linewidth]{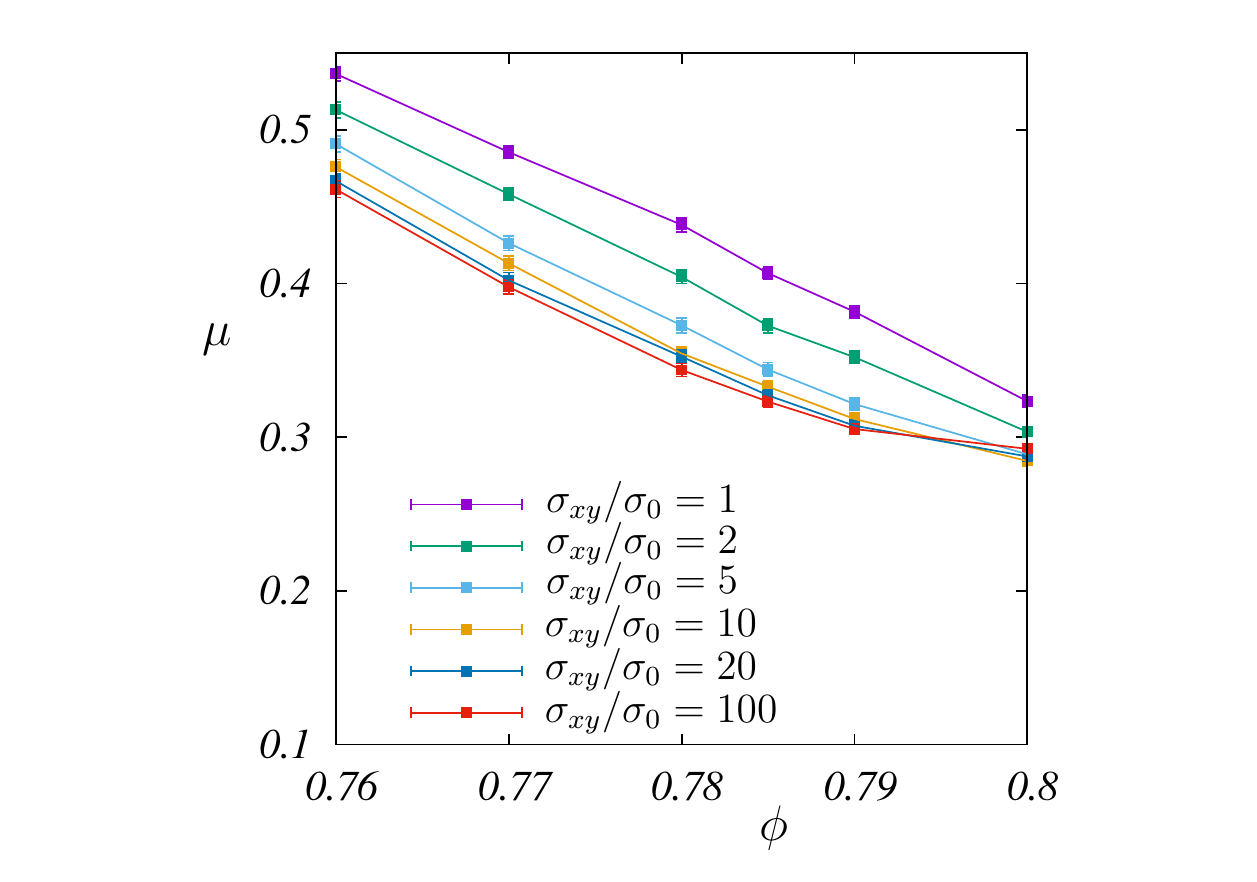}

\caption{ (Color online) Observed stress anisotropy $\mu$ from the simulation of suspensions.  The values of $\mu$ calculated from the theory (Fig. 4 in the main text) are in semiquantitative  agreement with these results. However,  the simulations show a larger range of variation.
} 
\label{mu_figure}
\end{figure}

\clearpage

\section*{Clustering in Force Space}

As observed from the pair correlation functions in height space (Fig. 2 in the main text), there is a clustering of the height vertices as the shear stress is increased.
To quantify this behaviour we analyze the radially averaged correlation function
\begin{equation}
g_2(h) = \frac{1}{2 \pi h}\int d^2 \vec{h} ~g_2(\vec{h}) ~\delta \left( h - |\vec{h}| \right).
\end{equation}
 This radial correlation function is fit well at small force scales by the following form
\begin{equation}
g_2(h) = 1 + \text{C} \left(\exp \left( \frac{1}{a+b h^2} \right) -1\right).
\label{g2h_fitting_form}
\end{equation}
As an example, we plot the fit using this form for $\phi = 0.76$ and $\sigma_{xy} = 10 \sigma_0$ in Fig. \ref{g2h_fit_figure}, showing that this form captures the behaviour at small force scales accurately.

\begin{figure}[h!]
\hspace*{-1.2cm}
\includegraphics[width=0.45\linewidth]{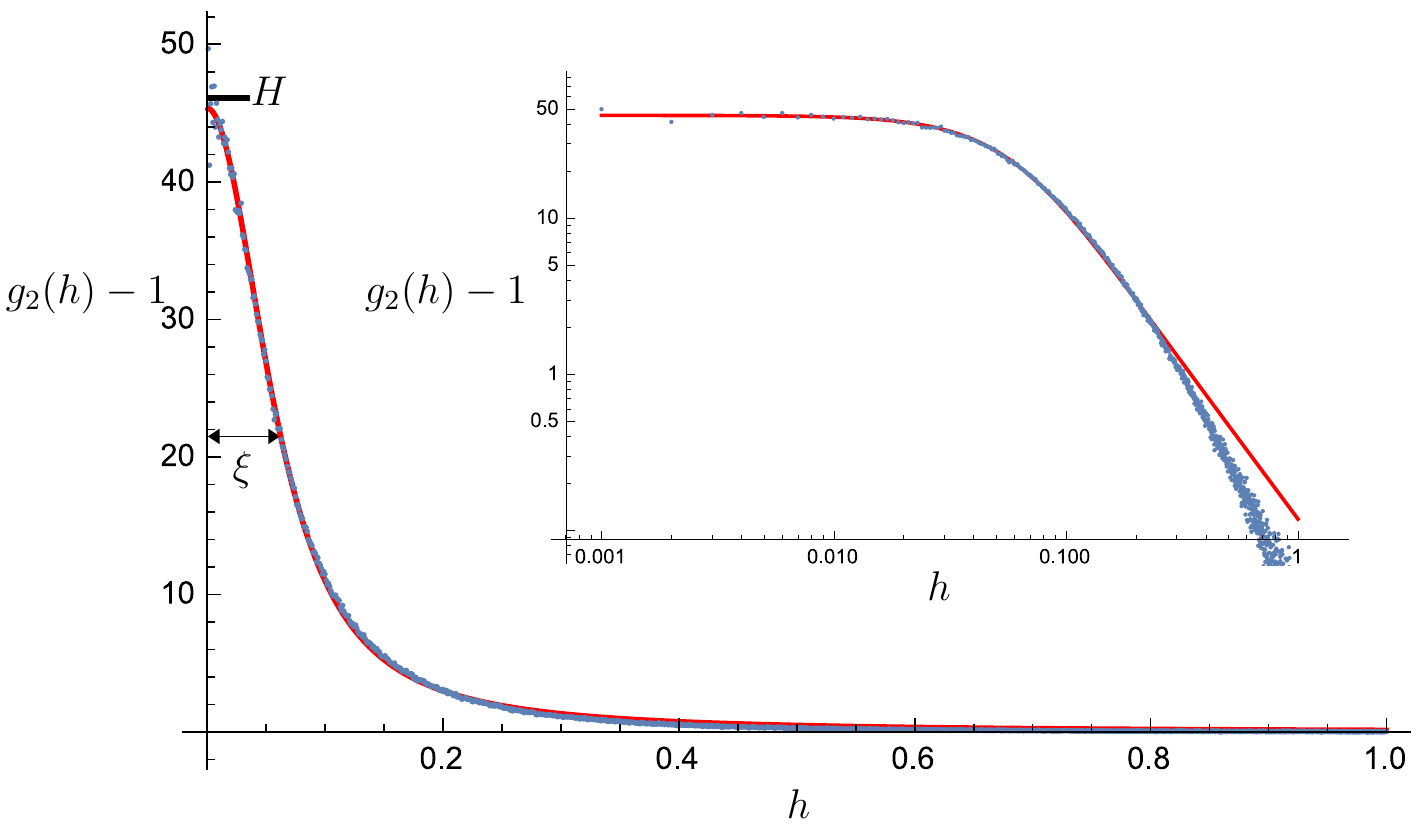}
\caption{ (Color online) Comparison of $g_2(h)$ for $\phi = 0.76$, $\sigma_{xy} = 10 \sigma_0$ (dots), and the fit (solid line) using the form given in Eq. (\ref{g2h_fitting_form}), with  $C = 5.17576, a =  0.438877$ and  $b = 43.8752$. (Inset) the same data in a log-log plot.
} 
\label{g2h_fit_figure}
\end{figure}

Using this fit, we compute three quantities that provide information about the clustering at small force scales, 
We compute

\begin{itemize}

\item The peak height $H$, given by
\begin{equation}
H = g_2(0) - 1 =C \left(e^{\frac{1}{a}}-1\right).
\end{equation}

\item The clustering length scale $\xi$ defined as the full width at half maximum of $g_2(h)$, given by
\begin{equation}
\xi =  \frac{\sqrt{1-a \log \left(\frac{1}{2} \left(e^{\frac{1}{a}}+1\right)\right)}}{\sqrt{ b \log
   \left(\frac{1}{2} \left(e^{\frac{1}{a}}+1\right)\right)}}.
\end{equation}
In the theory developed in the main text, we do not consider the region within $\xi$, which corresponds to very small forces in our statistical mechanics model.

\item The clustering intensity defined as the area
\begin{equation}
\mathcal{I} = \int_{0}^{\xi} g_2(h) dh.
\end{equation}

\end{itemize}
\begin{figure}[h!]
\hspace*{-1.2cm}
\includegraphics[width=0.45\linewidth]{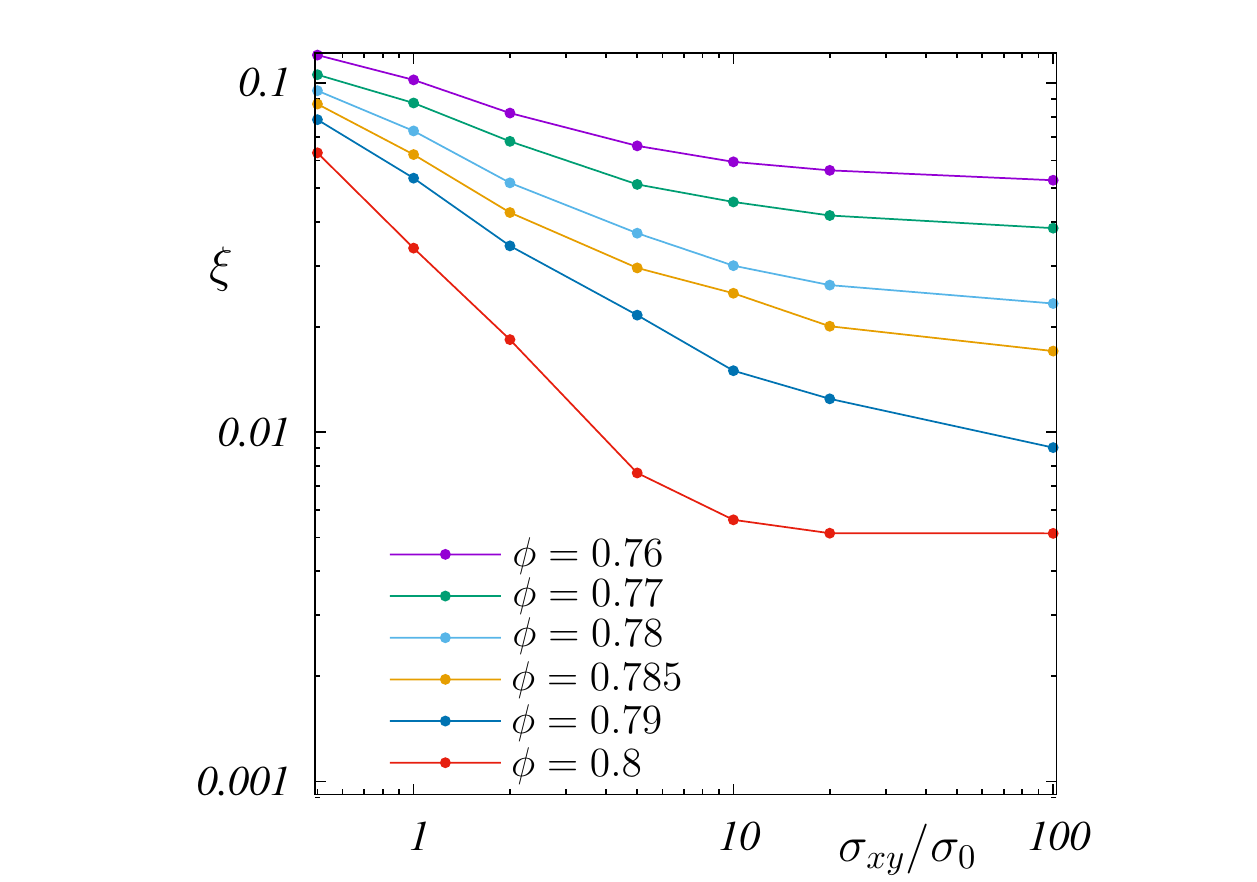}
\hspace*{-1.2cm}
\includegraphics[width=0.45\linewidth]{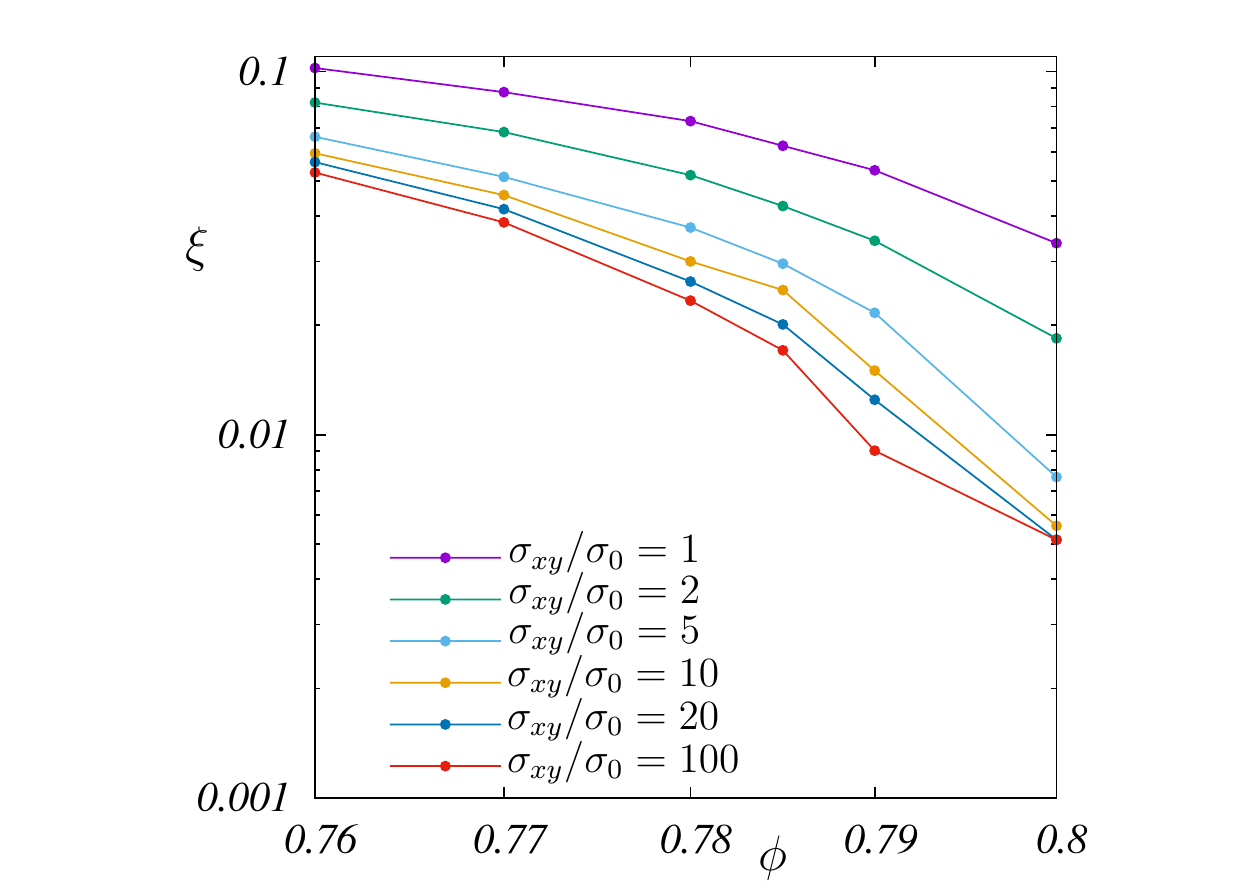}

\caption{ (Color online) Observed clustering length scale $\xi$ from the data. These are much smaller than the scales relevant in the effective theory discussed in the main text, where the typical force scales are $\sim 1$.} 
\label{Lengthscale_figure}
\end{figure}

\begin{figure}[h!]
\hspace*{-1.2cm}
\includegraphics[width=0.45\linewidth]{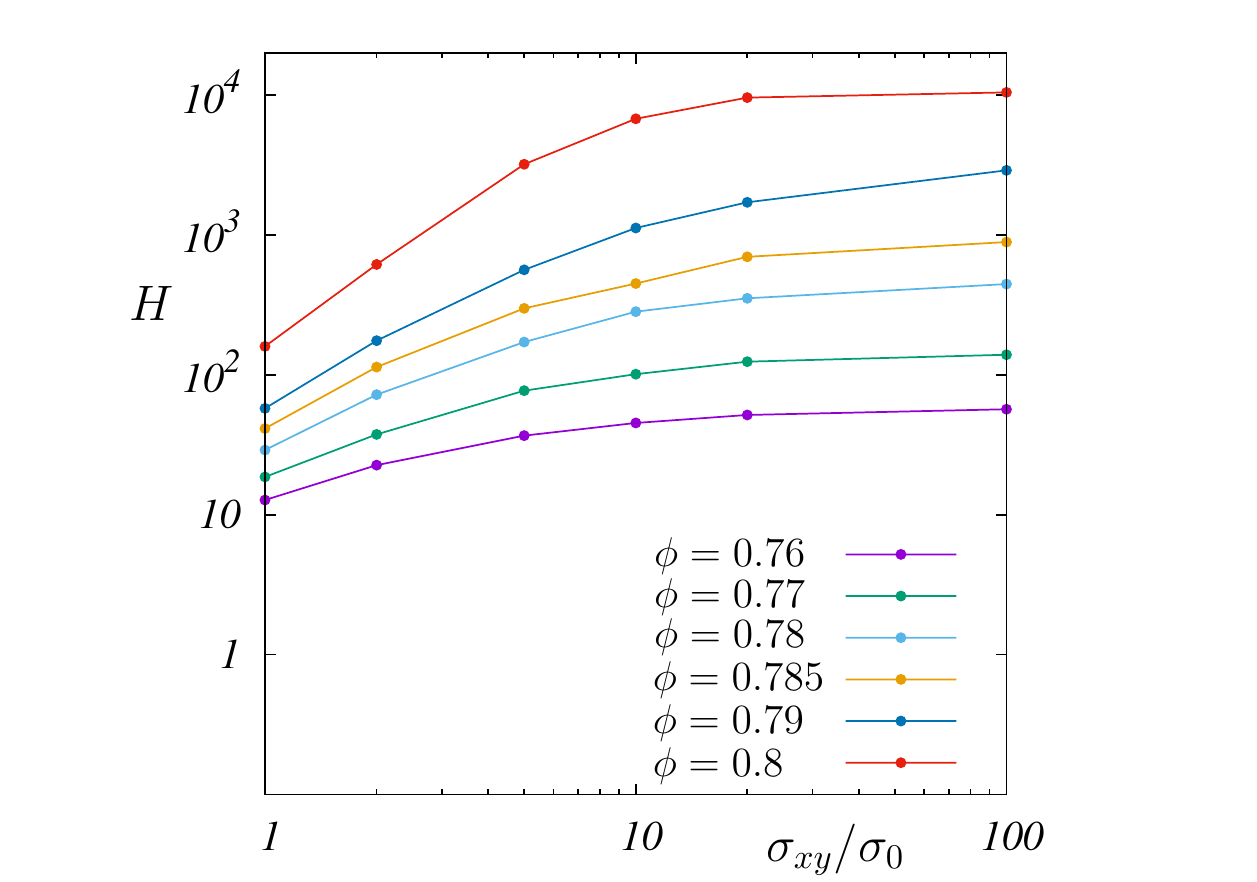}
\hspace*{-1.2cm}
\includegraphics[width=0.45\linewidth]{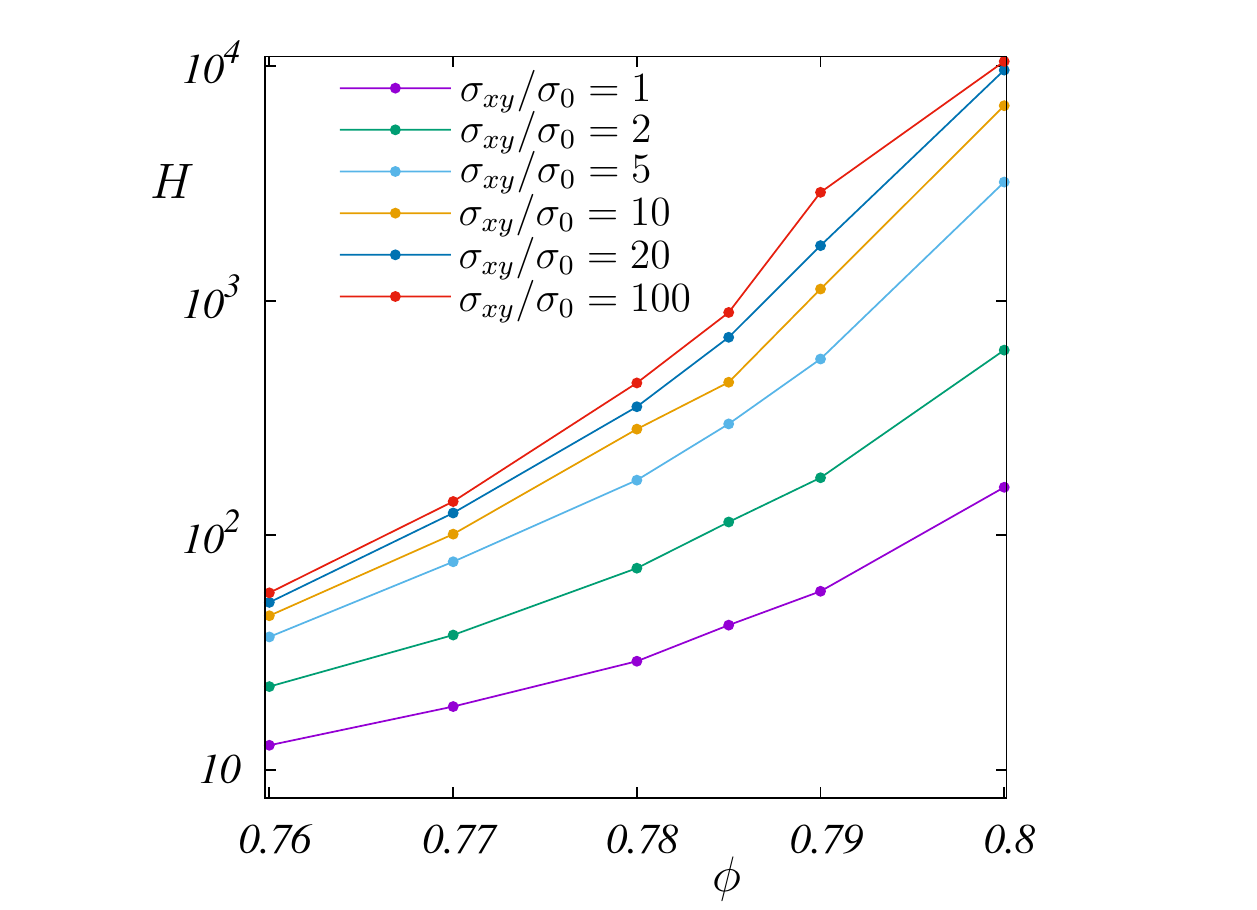}

\caption{ (Color online) Observed height $H$, of the peak in $g_2(h)$.} 
\label{Peak_height_figure}
\end{figure}
\begin{figure}[h!]
\hspace*{-1.2cm}
\includegraphics[width=0.45\linewidth]{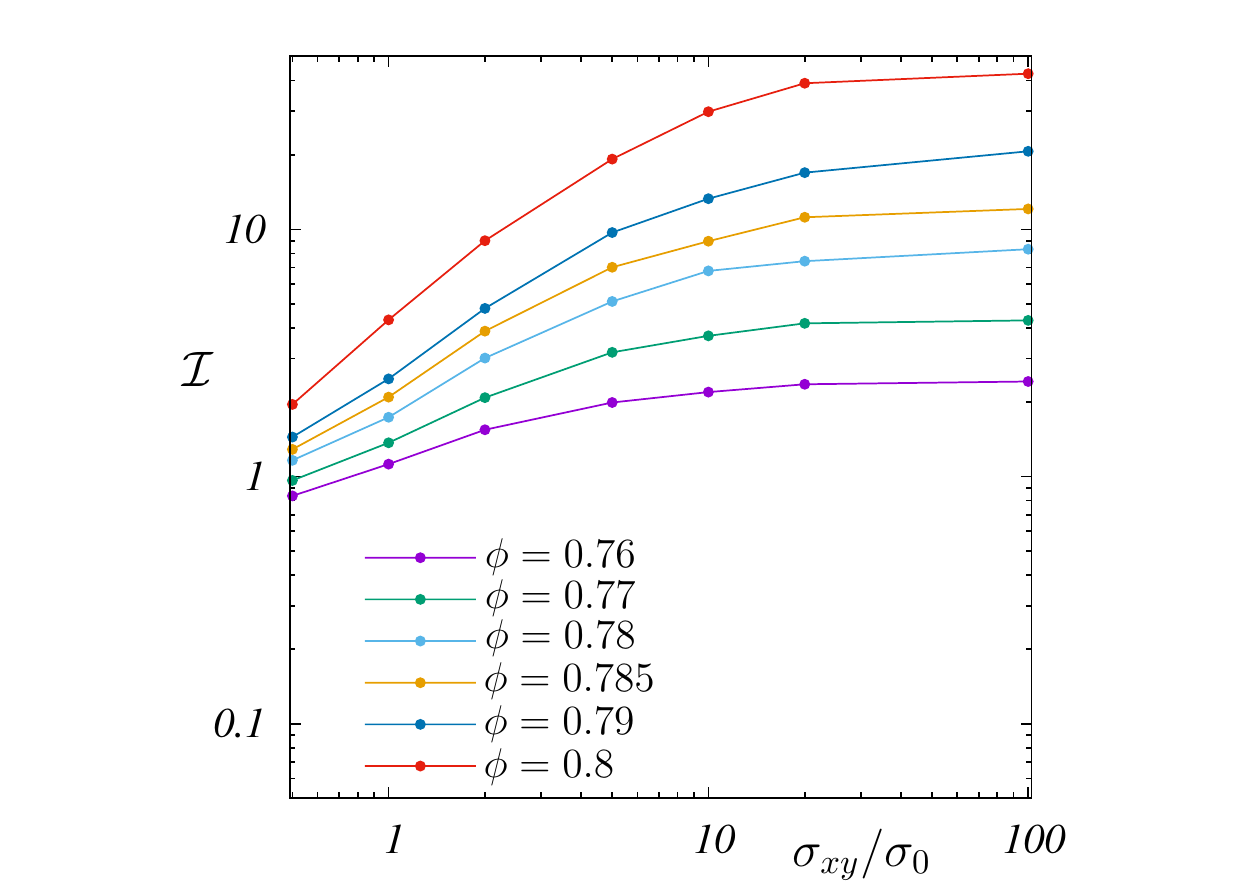}
\hspace*{-1.2cm}
\includegraphics[width=0.45\linewidth]{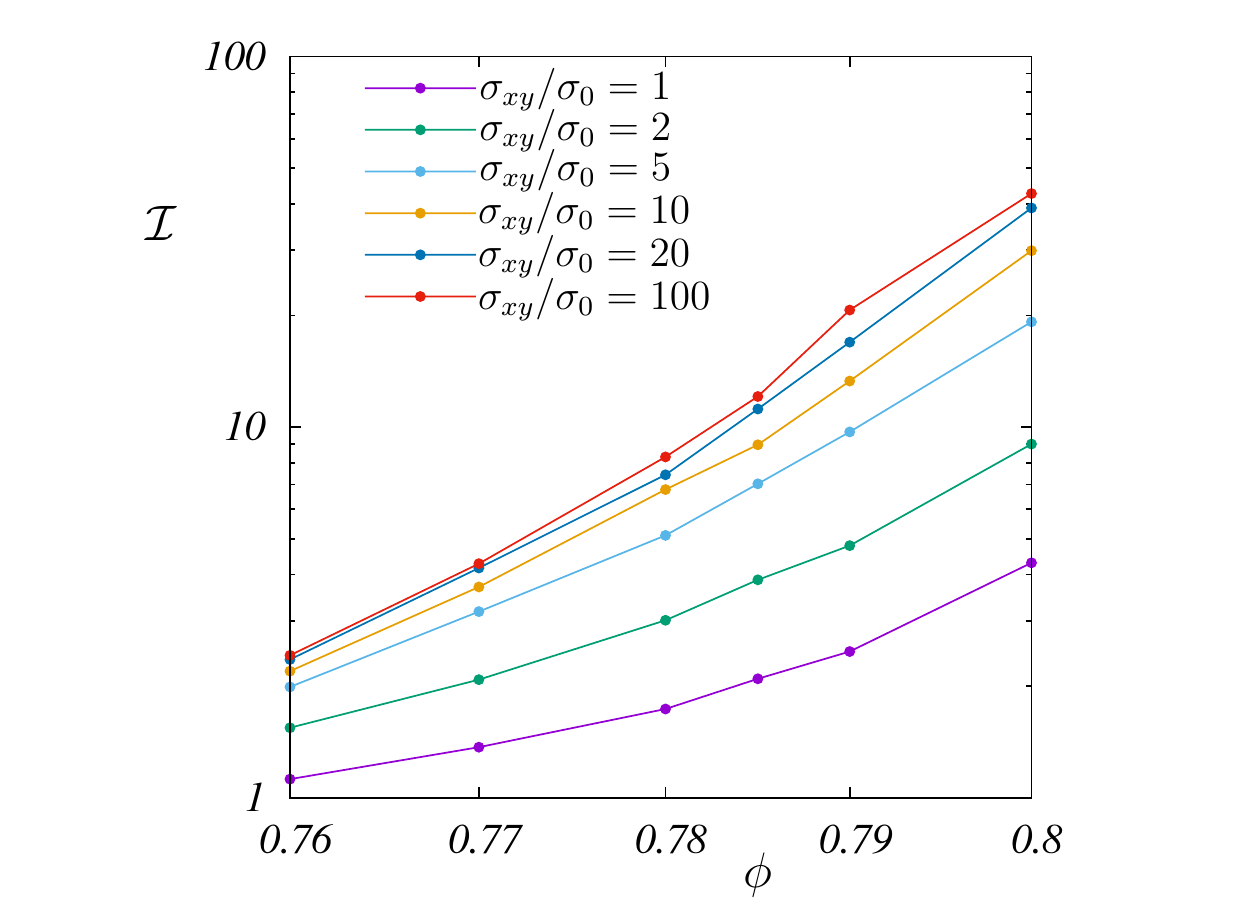}

\caption{ (Color online) Observed clustering intensity $\mathcal{I}$ from the data.
} 
\label{Intensity_figure}
\end{figure}

\clearpage

\subsection{Rotation of Pair Correlation Patterns}

As shown in the main text, the ``lobes'' of $g_2(h_x,h_y)$ representing regions where the correlations are higher than that of an ideal gas, rotate as $\phi$ is increased.  We quantify this rotation by analyzing the lobes in $g_2(h_x,h_y)$. As an example the Pair Correlation Function of Vertices (PCFV) for $\phi = 0.77$ and $\sigma_{xy} = 1 \sigma_0$ is shown in Fig. \ref{sq_explanatory_figure}. This displays a characteristic ``butterfly" pattern, with four lobes. The angles $\theta_1$ and $\theta_2$, defined in Fig. \ref{sq_explanatory_figure}, show a clear evolution with both $\phi$ and $\sigma_{xy}$, as shown in Fig. \ref{sq_figure}.
\begin{figure}[h!]
\includegraphics[height=0.35\linewidth]{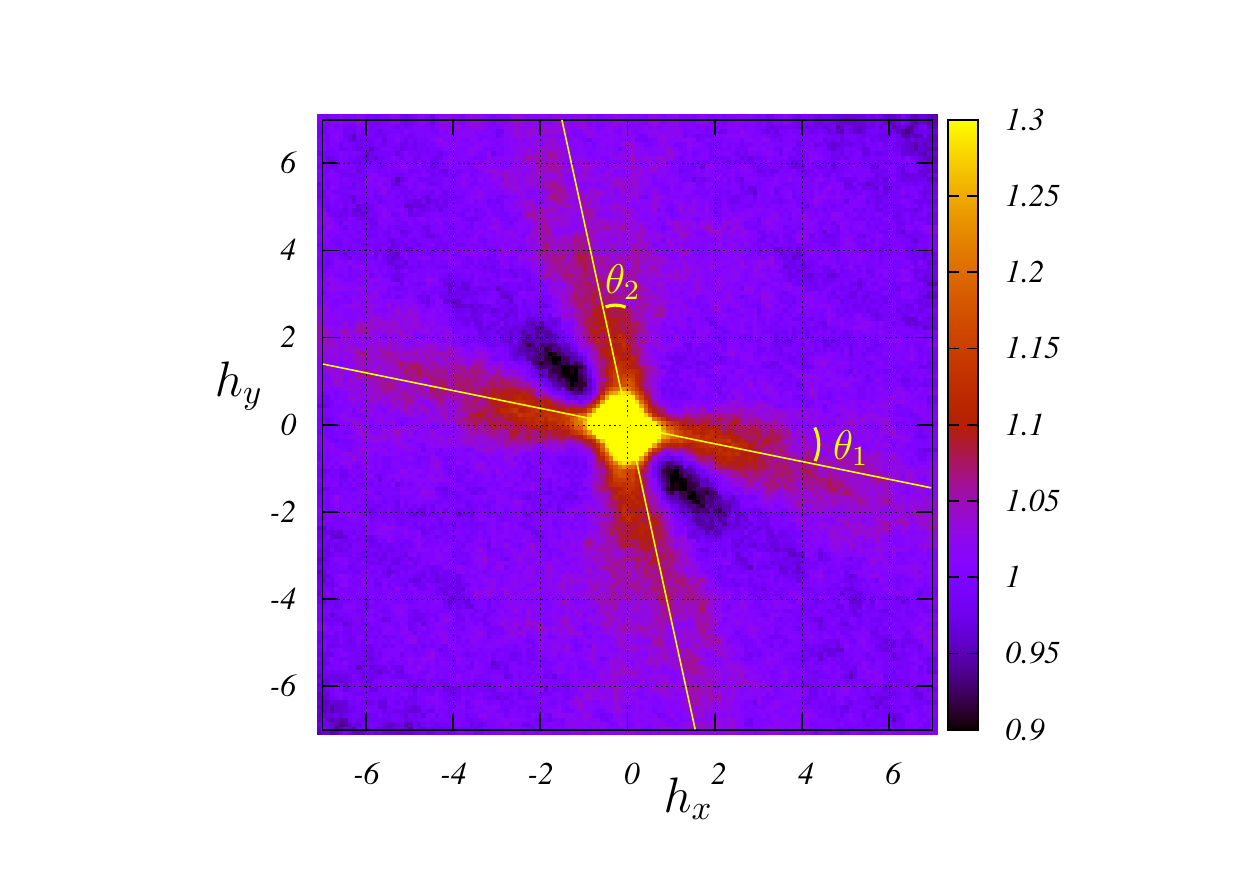}
\caption{ (Color online) Observed Pair Correlation Function of Vertices (PCFV) $g_2(\vec{h})$ at $\phi = 0.77, \sigma_{xy} = 1 \sigma_0$. We use the angles $\theta_1$ and $\theta_2$ to quantify the change in anisotropy as $\phi$ and $\sigma_{xy}$ are varied.} 
\label{sq_explanatory_figure}
\end{figure}
\begin{figure}[h!]
\includegraphics[height=0.3\linewidth]{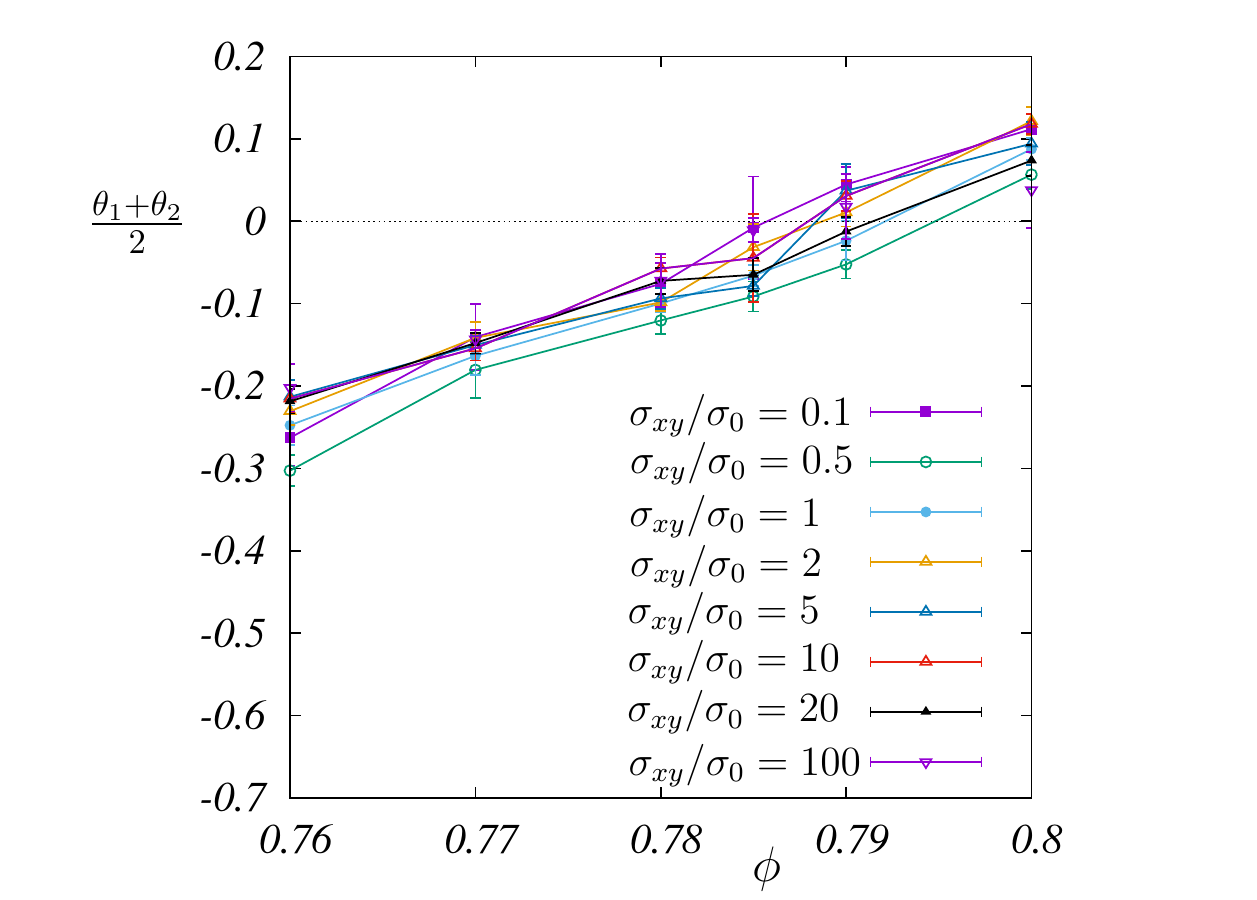}
\includegraphics[height=0.3\linewidth]{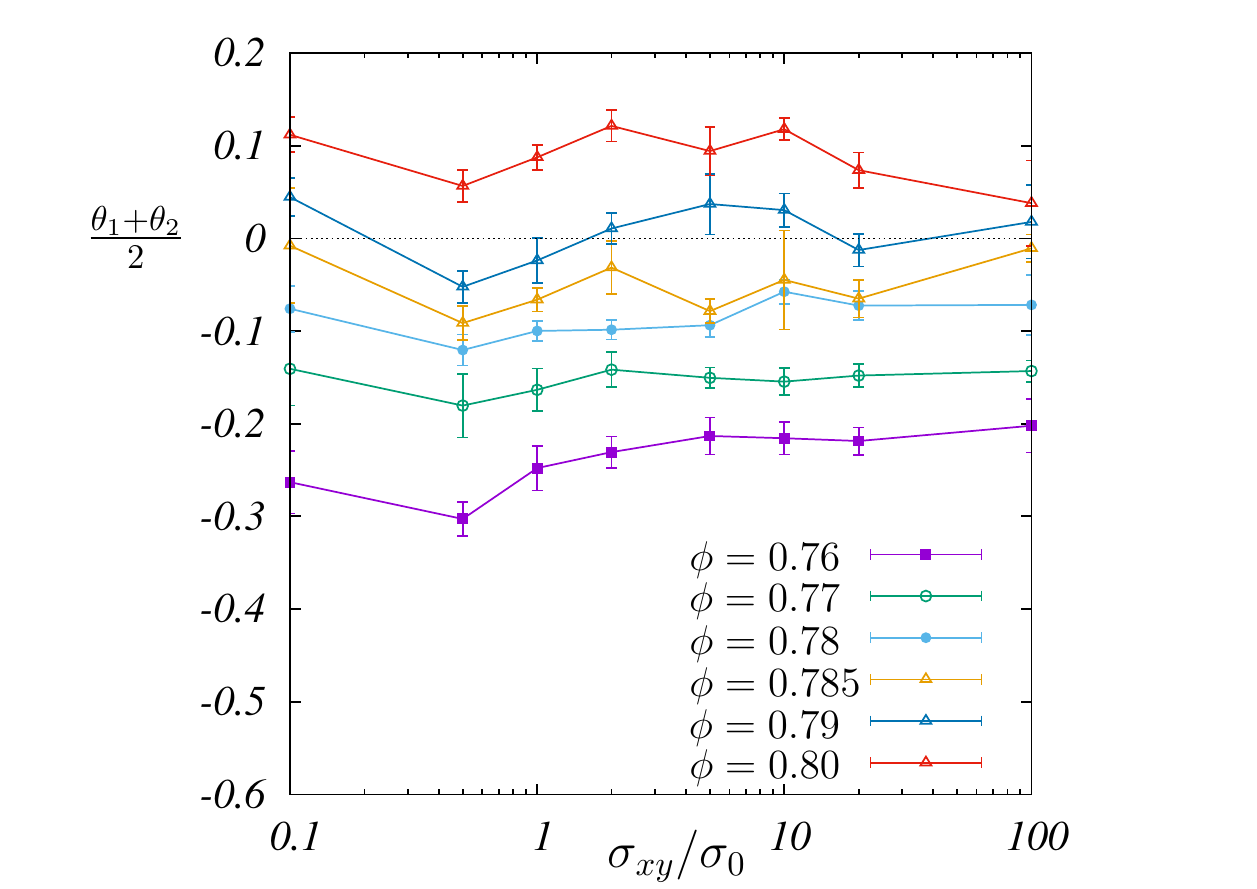}
\caption{(Color online) Observed rotation in the lobes (defined in Fig. \ref{sq_explanatory_figure}) of the Pair Correlation Function of Vertices (PCFV) $g_2(\vec{h})$.} 
\label{sq_figure}
\end{figure}

\vspace{5cm}
\subsection{Results for Viscosity}
\begin{figure}[h!]
\includegraphics[width=0.45\linewidth]{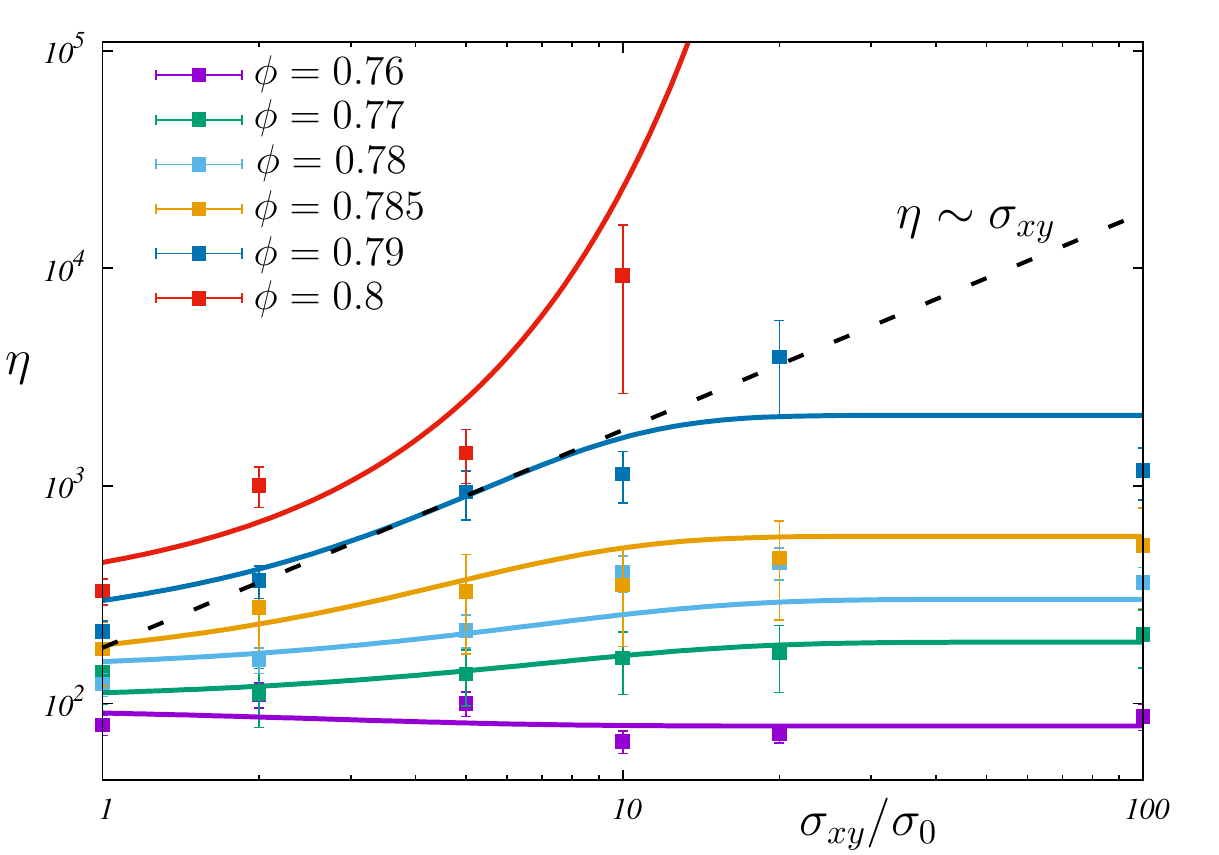}
\caption{(Color online) Viscosity ($\eta$) as a function of the imposed shear stresses ($\sigma_{xy}$), computed using Eq. (\ref{eta_mu_SI}), at different packing fractions ($\phi$). The points at which $\eta \sim \sigma_{xy}$ (dashed line) define the limits of the DST regime. }
\label{viscosity_figure}
\end{figure}


\subsection{Monte Carlo Sampling of the Energy function}

We treat the system using the NPT ensemble \cite{panagiotopoulos1987direct}, allowing for fluctuations in box shape \cite{yashonath1985monte}.
We fix $\Gamma _{\text{yy}} = -\Gamma _{\text{xx}} = \sigma = 15$
as observed from the data.
We also fix the magnitude of $\Gamma _{\text{xy}}$ and $\Gamma _{\text{yx}}$ to be equal, since $N_1 \approx 0$ as observed from simulations (see Fig. \ref{n1_figure}). The shape, and the area, of the force tiling is then determined by a single shape parameter $\mu$.

While performing Monte Carlo simulations of the interacting gas of height vertices, it becomes necessary to avoid clustering of the vertices as the density is increased. Therefore in addition to the potential given in Eq. (5) in the main text, we add a very short ranged ``hard-core" potential that prevents vertices from approaching very close to each other. We choose this hard-core potential to be a smoothly varying function of the form
\begin{equation}
V_{2,HC}(\vec{h}) = \exp((h_{HC}/|h|)^2) -1,
\end{equation}
where we choose $h_{HC} = 0.02$, much smaller than the intermediate force scales $\approx 1$ which is the focus of our study.  
Finally, in order to avoid long-range effects which are sensitive to numerical error induced by the low statistics of $g_2(\vec{h})$ at large force scales, we cut off the potential at a distance beyond which the anisotropy becomes unimportant. This is done by multiplying the potential with a Fermi function that falls off sharply at a distance $h_{CO} = 10$. We have
\begin{equation}
V_{\phi,\sigma}(\vec{h}) = \frac{1}{1 + \exp\left(3(|h| - h_{CO})\right)}\left(V_{2}(\vec{h}) + V_{2,HC}(\vec{h})\right),
\end{equation}
Finally we use this potential $V_{\phi,\sigma}(\vec{h})$ to perform Monte Carlo simulations of the interacting gas of vertices using the Metropolis algorithm (and $\beta = 1$). The displacement of each vertex is chosen from a Gaussian distribution with variance $10^{-4}$, and periodic boundary conditions are imposed using the dimensions of the force tiling box $(\vec{\Gamma}_x, \vec{\Gamma}_y)$. We also perform changes to the dimensions of the force tiling box, with the vertices being transformed affinely with every global change of the box shape. We attempt a change in the dimensions of the box at every tenth Monte Carlo step, with weights chosen using the energy 
\begin{equation}
E \equiv \sum_{i \ne j} V_{\phi,\sigma}(\vec{h}_i - \vec{h}_j) + N_v f_p^* A.
\end{equation}
We use these simulations to verify that the pair correlations generated using these potentials match the original $g_2(\vec{h})$ obtained from the NESS of simulated suspensions (as shown in Fig. 2 of the main text).

Next, in order to compute the ``free energy''  function $\mathcal{F}_{\mu;\phi, \sigma}$ of the system, we sample the ``energy'' function $\epsilon_{\phi,\sigma}(\mu,N_{v})$ given in Eq. (6) in the main text. 
We perform this sampling as follows.
For every realization of the system at a different $\mu$ (which defines the shape and the size of the confining box), we make the affine transformation
\begin{eqnarray}
 \left({\begin{array}{c} h_x \\ h_y \end{array}}\right) = \left({\begin{array}{cc}  \Gamma_{xx} & \Gamma_{xy}\\ \Gamma_{yx} & \Gamma_{yy} \end{array}}\right) \left({\begin{array}{c} s_x \\ s_y \end{array}}\right),
\end{eqnarray}
where the positions $\vec{s}_i$ are now confined to be within a $1 \times 1$ box.
In terms of the scaled coordinates $\{\vec{s}_i\}$, we have
\begin{eqnarray}
\exp(-\epsilon_{\phi,\sigma}(\mu,N_v)) =  \int_{1 \times 1} \prod_{i = 1}^{N_v} d \vec{s}_i \exp \left( - \sum_{i,j} \tilde{V}_{\phi,\sigma}(\vec{s}_i - \vec{s_j}) \right),
\label{free_energy_equation_supplemental}
\end{eqnarray}
where $\tilde{V}$ is now the affinely transformed potential. We perform a Monte Carlo (MC) sampling to obtain $\epsilon_{\phi,\sigma}(\mu,N_v)$ for different values of the number of vertices $N_v = 128, 256, 512, 768$ and $1024$. 

For a fixed $\mu$, we create an ensemble of configurations $C_n \equiv \{\vec{s}_{i}^{n}\}$ with $n = 1,2...N_{MC}$ with positions chosen uniformly within the $1 \times 1$ box. 
The computational cost of arranging $N_{v}$ points in the box and computing $\left( \sum_{i,j}\tilde{V}_{\phi,\sigma}(\vec{s}_{i}^{n}-\vec{s}_{j}^{n}) \right)$ for each configuration is  $O(N_{v}^{2})$. For $N_v \simeq 3000$ points, which is the actual number of vertices observed in the force tiles from the NESS, this would require  $10^{6}$ moves at each configuration, making the simulation prohibitively expensive. Therefore, we used the $\epsilon_{\phi,\sigma}(\mu, N_{v})$ computation for smaller sizes ($N_{v}=512$ and $1024$) to extrapolate to $N_v = 3000$. To perform this extrapolation, we used the data at smaller values of $N_v$ to find a scaling form. We find a reasonably good scaling collapse with the following scaling form
\begin{equation}
\epsilon_{\phi,\sigma}(\mu,N_v) = N_v^3 e_{\phi,\sigma}(\mu),
\end{equation}
where the function $e_{\phi,\sigma}(\mu)$ is a universal scaling function that is independent of $N_v$ (for large $N_v$). As shown in  Fig. \ref{NvScaling}, this $N_v^3$ scaling works well for larger $N_v$.  

\begin{figure}[h!]
\hspace*{-0.2cm}
\includegraphics[width=0.55\linewidth]{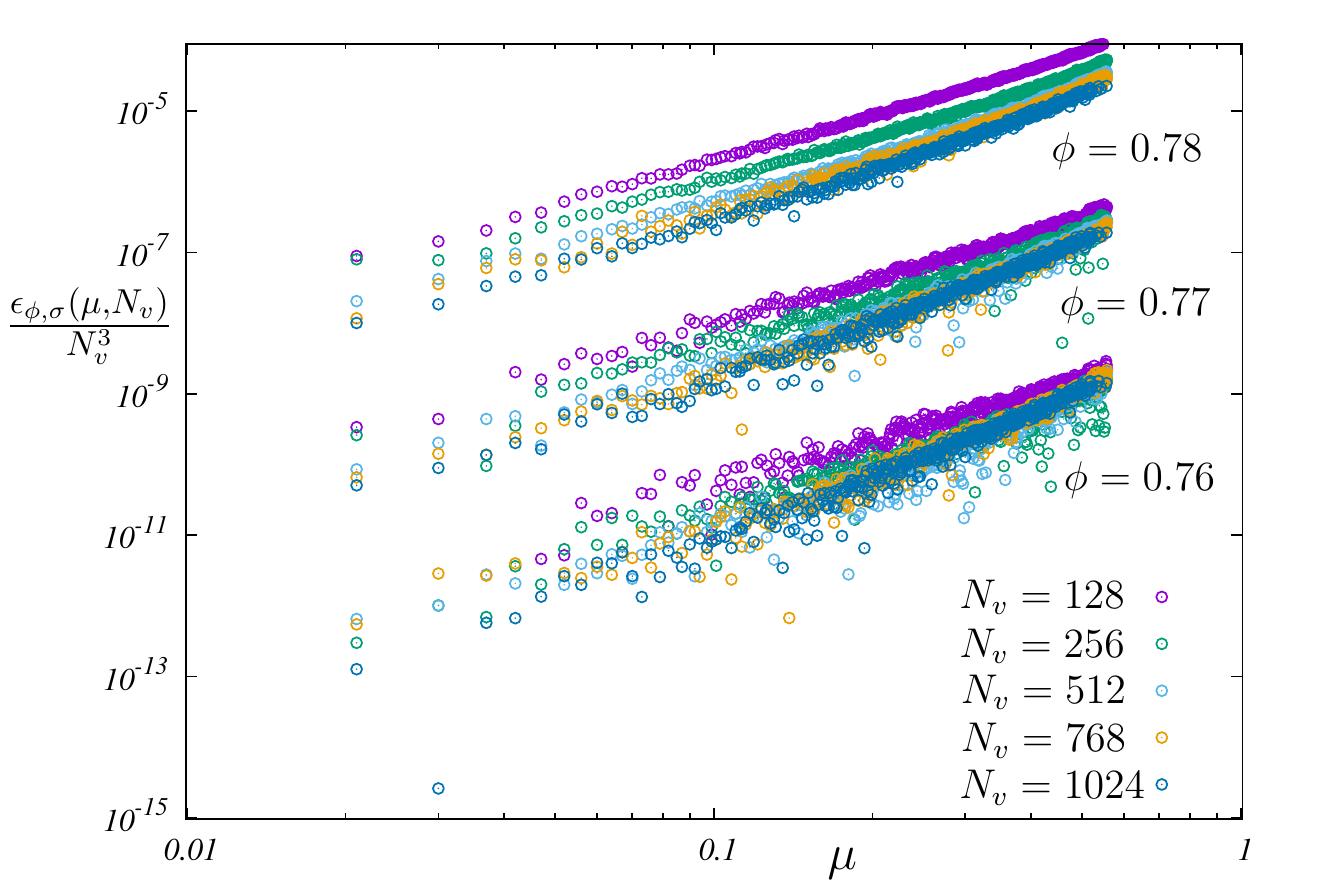}
\caption{(Color online) The scaling collapse of  $\epsilon_{\phi,\sigma}(\mu, N_{v})$  for different values of $N_v$:  $N_v = 128, 256, 512, 768, 1024$, at different values of $\phi = 0.76, 0.77$ and $0.78$ with $\sigma_{xy}$ held fixed at $100 \sigma_0$. The curves for different $\phi$ have been shifted by two decades to aid visualization. We find that a reasonably good scaling collapse emerges with increasing $N_v$.}
\label{NvScaling}
\end{figure}

The number of MC steps, $N_{MC}$, ranged from $25000$ for $N_v=1024$ to $50000$ for $N_v=512$. Using these configurations, we computed $\exp \left( -\sum_{i,j}\tilde{V}_{\phi,\sigma}( \vec{s}_{i}^n-\vec{s}_{j}^n ) \right)$, which we used to calculate $\epsilon_{\phi,\sigma}(\mu,N_v)$ by averaging as follows:
%
%
\begin{eqnarray}
\epsilon_{\phi,\sigma}(\mu,N_v) = -\log \left( \sum_{n=1}^{N_{MC}} \exp \left(-\sum_{i,j}\tilde{V}_{\phi,\sigma}( \vec{s}_{i}^{n}-\vec{s}_{j}^{n} ) \right) \Big{/}N_{MC} \right).
\label{free_energy_equation_supplemental2}
\end{eqnarray}
A typical series for $E^{n}_{\phi,\sigma}(\mu,N_{v}) = \sum_{i,j}\tilde{V}_{\phi,\sigma}( \vec{s}_{i}^n-\vec{s}_{j}^n )$, is shown in Fig. \ref{equilibration} (a) for $\phi=0.79$, $\sigma_{xy}= 5 \sigma_0$, $\mu=0.33$, and $N_v = 512$.  We also demonstrate that the  function $\epsilon_{\phi,\sigma}(\mu,N_v)$ asymptotes to an invariant form for $N_{MC} \simeq 20000$ by computing $\frac{\int \epsilon_{\phi,\sigma}(\mu, N_{v}) d\mu}{\int d\mu}$ for increasing $N_{MC}$ as shown in Fig. \ref{equilibration} (b).
%
\begin{figure}[h!]

\hspace*{-0.1cm}
(a)
\includegraphics[width= 0.45\linewidth]{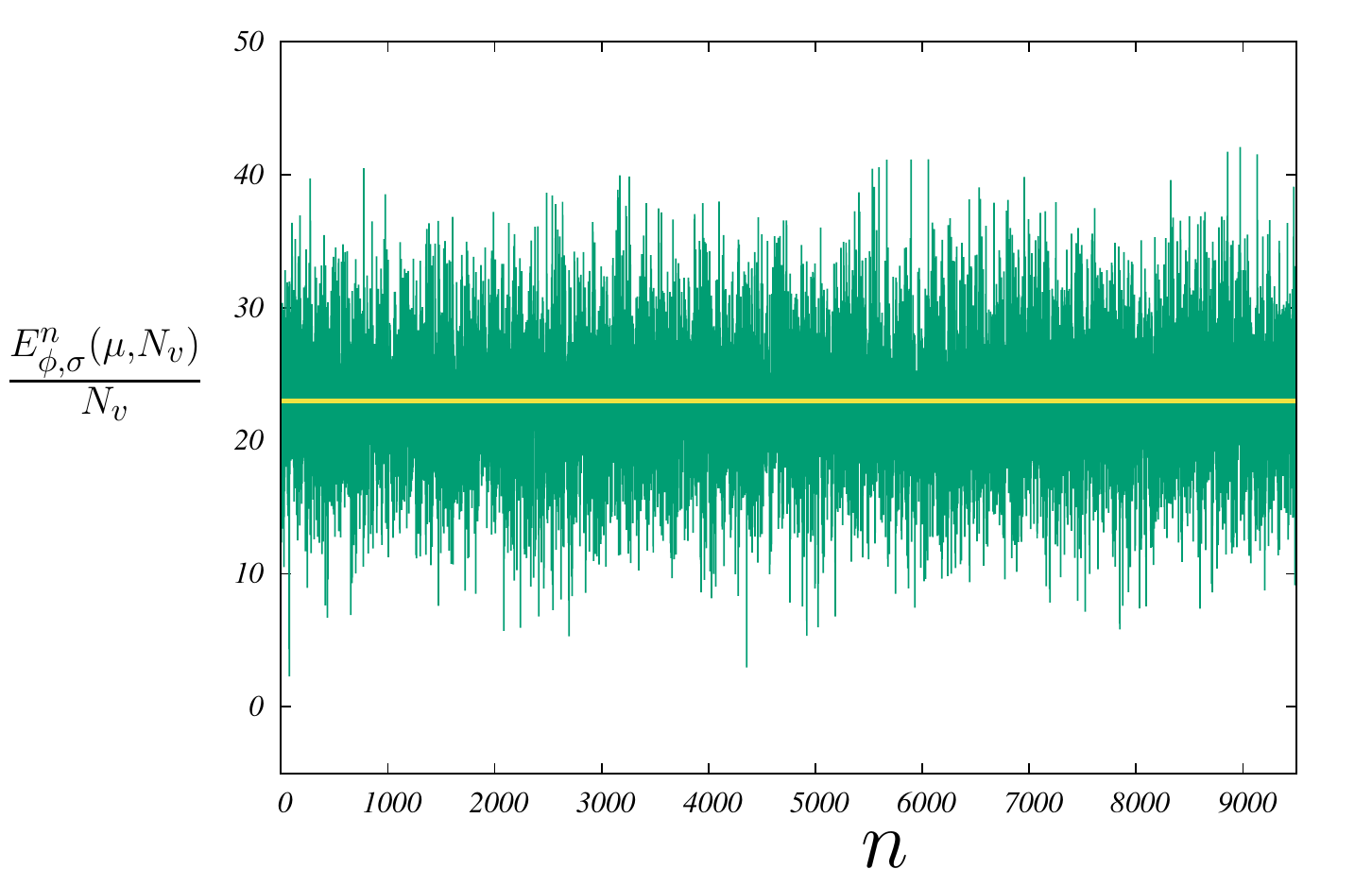}
\hspace*{-0.1cm}
(b)
\includegraphics[width=0.45 \linewidth]{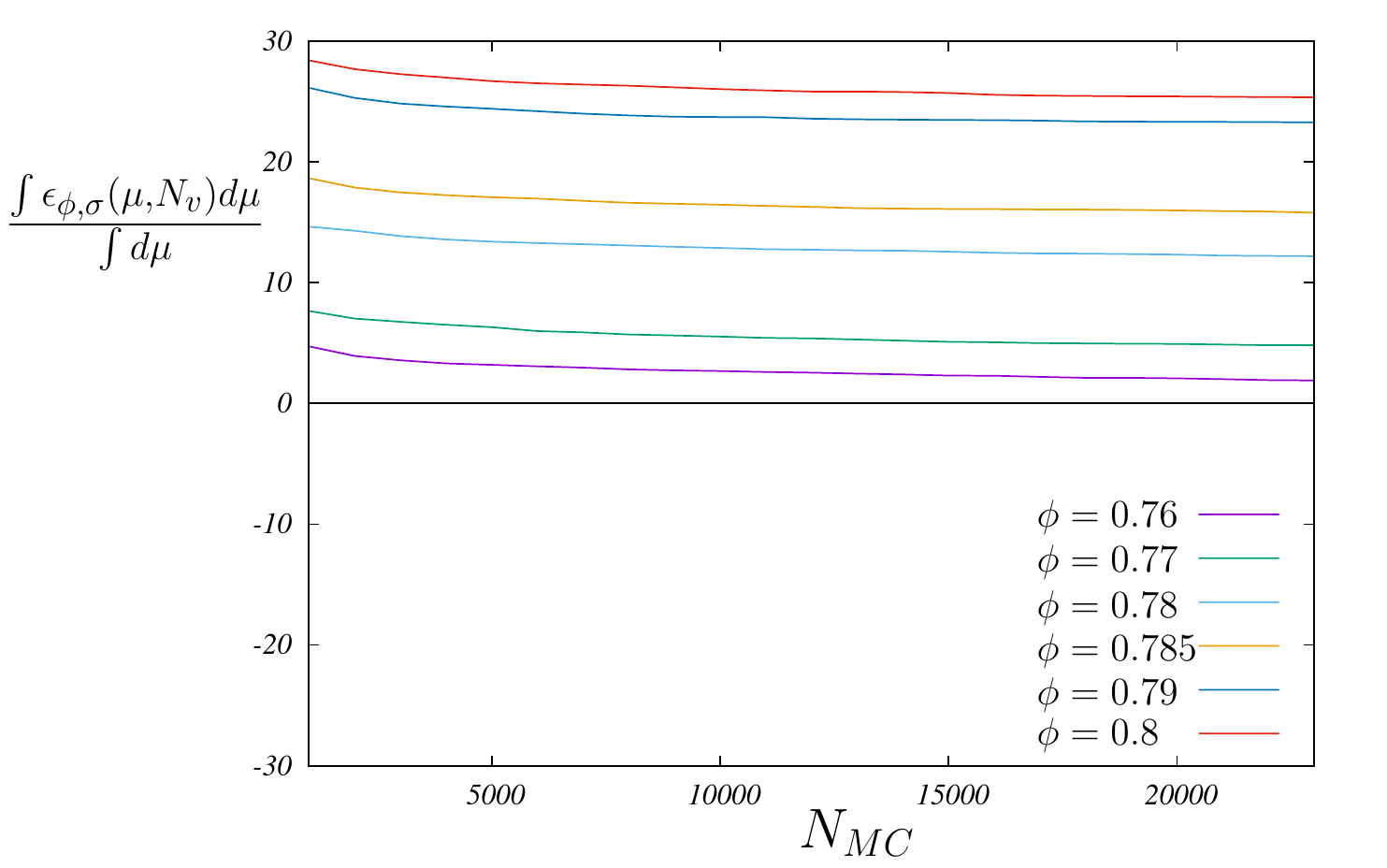}
\caption{(Color online) (a) The average ``energy'' per vertex of each configuration, $\frac{E^{n}_{\phi,\sigma}(\mu = 0.33, N_{v})}{N_{v}}$, at $\phi = 0.79$ and $\sigma = 5 \sigma_0$ plotted for different configurations $n = 1,2 ... N_{MC}$. (b) The evolution of $\frac{\int \epsilon_{\phi,\sigma}(\mu, N_{v}) d\mu}{\int d\mu}$ with $N_{MC}$ for different potentials with varying $\phi$. We find that this  asymptotes to an invariant form for $N_{MC} \simeq 20000$.}
\label{equilibration}
\end{figure}

\clearpage

\end{appendix}

\end{widetext}

\end{document}